\documentclass[twocolumn]{aastex631}
\graphicspath{{./}{figures/}}

\received{\today}
\submitjournal{AJ}

\shorttitle{Modelling Photometric Completeness}
\shortauthors{Harris \& Speagle}
\usepackage{longtable}
\usepackage{threeparttable}
\usepackage{bm}
\usepackage{mathrsfs}
\usepackage{amsmath}
\usepackage{CJKutf8}
\usepackage{todonotes}
\usepackage{hyperref}

\DeclareMathOperator*{\argmax}{argmax}
\DeclareMathOperator*{\argmin}{argmin}

\usepackage{bbm}
\usepackage{color}
\usepackage{fancyvrb}

\DefineVerbatimEnvironment{Highlighting}{Verbatim}{commandchars=\\\{\}}
\usepackage{framed}
\definecolor{shadecolor}{RGB}{248,248,248}

\begin{document}

\title{Photometric Completeness Modelled With Neural Networks}

\correspondingauthor{William E. Harris}
\author[0000-0001-8762-5772]{William E. Harris}
\email{harris@physics.mcmaster.ca}
\affil{Department of Physics and Astronomy, McMaster University, Hamilton, ON L8S 4M1, Canada}

\author[0000-0003-2573-9832]{Joshua S. Speagle (\begin{CJK*}{UTF8}{gbsn}沈佳士\ignorespacesafterend\end{CJK*})}
\affiliation{Department of Statistical Sciences, University of Toronto, 9th Floor, Ontario Power Building, 700 University Ave, Toronto, ON M5G 1Z5, Canada}
\affiliation{David A. Dunlap Department of Astronomy \& Astrophysics, University of Toronto, 50 St George Street, Toronto ON M5S 3H4, Canada}
\affiliation{Dunlap Institute for Astronomy \& Astrophysics, University of Toronto, 50 St George Street, Toronto, ON M5S 3H4, Canada}
\affiliation{Data Sciences Institute, University of Toronto, 17th Floor, Ontario Power Building, 700 University Ave, Toronto, ON M5G 1Z5, Canada}

\begin{abstract}
In almost any study involving optical/NIR photometry, understanding the completeness of detection and recovery is an essential part of the work.  The recovery fraction is, in general, a function of several variables including magnitude, color, background sky noise, and crowding.  We explore how completeness can be modelled, {with the use of artificial-star tests,} in a way that includes all of these parameters \emph{simultaneously} within a neural network (NN) framework. The method is able to manage common issues including asymmetric completeness functions and the bilinear dependence of the detection limit on color index.  We test the method with two sample HST (Hubble Space Telescope) datasets:  the first involves photometry of the star cluster population around the giant Perseus galaxy NGC 1275, and the second involves the halo-star population in the nearby elliptical galaxy NGC 3377.  The NN-based method achieves a classification accuracy of $>$\,94\%, and produces results entirely consistent with more traditional techniques for determining completeness.  Additional advantages of the method are that none of the issues arising from binning of the data are present, and that a recovery probability can be assigned to every individual star in the real photometry. Our data, models, and code (called COINTOSS) can be found online on Zenodo at the following link: \url{https://doi.org/10.5281/zenodo.8306488}.
\end{abstract}


\section{Introduction} \label{intro}

In photometric studies it is often necessary to understand the completeness of the measurements, as a function of magnitude, crowding, or other properties of the images.  Completeness, defined in rough terms as the fraction $f$ of detected objects relative to the true number present in the image, can be evaluated quantitatively through artificial-star tests (ASTs).  The ASTs yield other key characteristics of the data including the internal measurement uncertainties in magnitude and position,  systematic measurement bias, and ``limiting magnitude'', which is normally taken to be the faintness level where the completeness has dropped to some fiducial level such as 50\% or 80\%.

At the basis of the completeness issue is an extremely simple question: in the first stage of the photometry, the image is searched for any objects brighter than a well defined detection threshold. At that stage, any object $i$ is either detected ($d_i=1$) or it is not ($d_i=0$).  That is, the detection process is a \emph{binary response function}. After the detection stage, various downstream selection cuts can then be applied to determine whether or not the objects are of high enough quality (e.g., have a signal-to-noise ratio above a specified threshold) to be included in a final photometric catalog, with objects that pass ($q_i=1$) being included and objects that do not ($q_i=0$) being excluded.  In the final catalog of objects, we define those with $y_i=q_i \times d_i = 1$ to be \textit{recovered}. 
Throughout the remainder of the following discussion, we will try to disambiguate between these notions of \emph{detection}, which only involve whether or not an object was identified at some stage of a photometric pipeline, versus \emph{recovery}, which measures whether or not an object made it into the final catalog after additional culling steps were applied.

Stating the problem this way raises the possibility that completeness may be suited for modelling by logistic regression (LR)\footnote{For detailed outlines of logistic regression in an astronomy context, see \citet{desouza+2015} and \citet{eadie+2022}.}.  Additional evidence for this view comes from the dependence of the \textit{recovery fraction} $f(m)$ on magnitude $m$, which typically behaves approximately as \citep{harris+2016}
\begin{equation}
f(m) = \frac{1}{1 + e^{\alpha (m - m_0)}} 
\label{eq:fcurve}
\end{equation}
where $m_0$ is the \emph{completeness limit} or ``limiting magnitude'' where $f(m)$ drops to 50\%, and $\alpha$ represents the steepness of decline of the curve around $m_0$.
This empirical interpolation formula is formally identical with the sigmoid function used in LR to describe the probability of recovery.
For example, \citet{rosolowsky+2021} use LR to model completeness for recovering extragalactic GMCs (Giant Molecular Clouds) in CO emission from PHANGS-ALMA imaging data. In their work, the recovery probability is correlated with cloud mass, surface intensity, and cloud virial parameters, all of which can be put into a single LR function.  

Completeness can therefore be viewed from two different statistical perspectives.  For an \emph{ensemble} of stars, the ratio $f(m)$ represents the fraction of objects present that are successfully recovered in a bin of width $dm$ at magnitude $m$.  For a \emph{single} star, we can instead think of this curve as representing $p(m)$, the probability of its recovery. While the distinction between these two perspectives may appear somewhat moot if both can be expressed by the same equation \citep{eadie+2022}, it is important to recognize that methods such as LR are ultimately trying to estimate $p(m)$, not $f(m)$, and that stellar recovery or non-recovery can ultimately be estimated (and used) on a star-by-star basis rather than a population average. In particular, we can explicitly relate the two by interpreting $f(m)$ as the long-run expectation of recovery $y \in \{0,1\}$ at a given $m$ with recovery probability $p(m)$ over many data realizations $n$.\footnote{Note that this also predicts that the associated standard deviation should be $\sigma_{f}(m,n) = \sqrt{p(m)(1-p(m))/n}$ \citep[cf.][]{bolte1989}.}

As will be seen below, however, LR in the simplest form of Eq.~\ref{eq:fcurve} may turn out not to be a good match to real data in important ways. The main causes for this shortcoming are biases in photometric measurements and two-band detection requirements, both of which lead to asymmetric and rapidly changing recovery criteria across the color-magnitude diagram that are difficult to model with simple LR methods. These aspects of the problem are discussed in more detail below.  As a result, we were led to explore  more complex extensions of LR that can reproduce the more complex behaviors seen in the data. We ultimately find that neural networks deliver excellent performance at predicting aggregate $f(m)$ values, along with individual recovery rates $y_i$, and are well motivated by the core concepts behind LR.

The aim of the present paper is to develop new formalism for photometric completeness in the optical/NIR regime, where the measurements of concern may be physically analogous to {quantities at other wavelengths} (luminosity, surface brightness, crowding/confusion etc.) but are parametrized very differently.  

The discussion is organized as follows:  In Section \ref{background}, some of the background of completeness studies in the literature is summarized.  Sections \ref{ngc1275} and \ref{ngc3377} present the photometry and artificial-star tests for two sets of data (imaging fields targeting the galaxies NGC 1275 and NGC 3377) that will be used as testbeds for development of the methodology. Section \ref{model_params} describes the underlying parameters that we consider in our modelling. In Section \ref{lrmodel}, we outline the basic LR model, highlight its shortcomings, and describe its extension into a neural network framework. In Section \ref{nn_ngc}, we demonstrate the excellent performance of the neural network models tested on the artificial star data and its application to the real data. We also discuss some of the potential benefits and shortcomings of our approach. Section \ref{summary} finishes with a quick summary. The code for the NN models, along with the data used to train them, are provided online on Zenodo at \url{https://doi.org/10.5281/zenodo.8306488}.

\begin{figure}
    \centering
    \includegraphics[width=.48\textwidth]{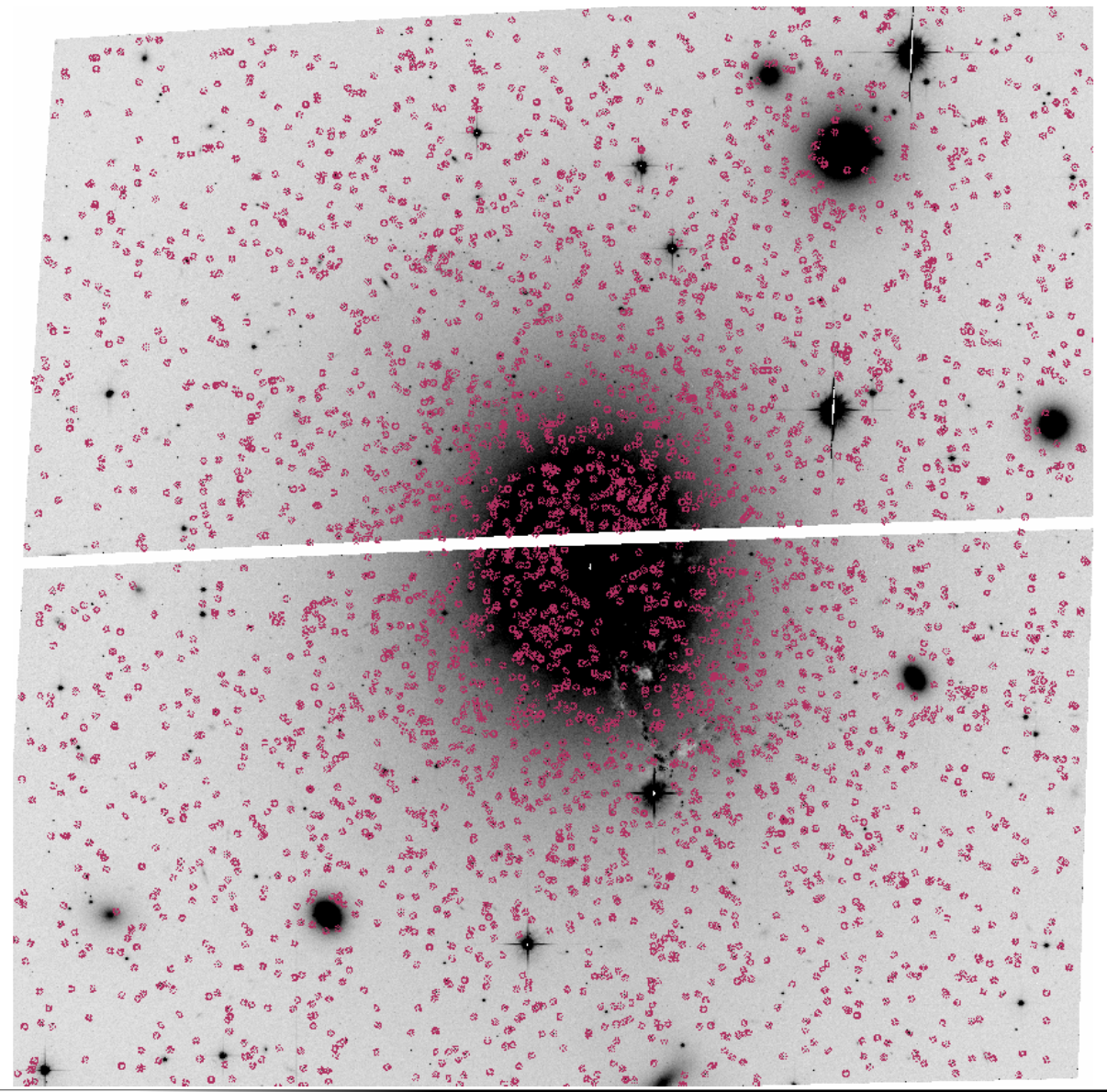}
    \caption{HST ACS image centered on the Perseus central giant NGC 1275. The field shown is $200''$ across, equivalent to 75 kpc at the distance of Perseus.  The maroon dots marks positions of star cluster candidates (${\rm F814W} < 25.5$, ${\rm F475W} - {\rm F814W} < 2.4$; see also the color-magnitude diagram in Fig. \ref{fig:cmd}).}
    \label{fig:ngc1275}
\end{figure}

\section{Some Background}\label{background}

There is now a vast literature of photometric studies in which completeness measurements are derived, and the present discussion can hope to mention only a few examples. Soon after the advent of CCD detectors for digital imaging, the procedures that became standard methodology for determining completeness were developed  \citep[e.g.][]{lupton_gunn1986,stetson1987,stetson1991,stetson_harris1988,drukier+1988,mateo1988,bolte1989,vallenari+1994}. Many of these studies particularly involved stellar populations in globular clusters, where a number of classic issues had to be confronted including highly nonuniform crowding levels and luminosity functions that rise steeply toward fainter magnitudes.

Before proceeding, the terms for completeness should be stated more precisely.  In a given magnitude bin $(m, m + dm)$, completeness $f$ as determined from the ASTs is normally defined as the number $n_{\rm rec}$ of recovered artificial stars measured in that bin, divided by the number $n_{\rm in}$ of input stars originally inserted into it.  This definition implicitly includes the effect of \emph{bin-jumping} (i.e. whenever a star inserted into a particular bin is detected and measured, but its measured magnitude is different enough to move it to a neighboring bin).  

In addition, $f$ can be plotted versus either \emph{input} magnitude or \emph{measured} magnitude \citep[see][]{stetson1991}.  If the mean measurement bias $\delta = \langle m_{\rm rec} - m_{\rm in} \rangle$ (i.e. the difference between the recovered and input magnitudes) is small, these two versions of $f(m_{\rm rec}) \approx f(m_{\rm in})$ will be similar.  However, for the real stars, the only available information consists of their measured magnitudes $m_{\rm rec}$.  In the following discussion, we will therefore drop the $m_{\rm rec}$ subscript and the conventions used here will be to define $f(m_i)$ as the number of measured stars found in the $i$th bin (regardless of their input magnitudes) relative to the number inserted into that bin, and to plot $f$ as a function of measured (= recovered) magnitude.

\begin{figure*}
    \centering
    \includegraphics[width=.48\textwidth]{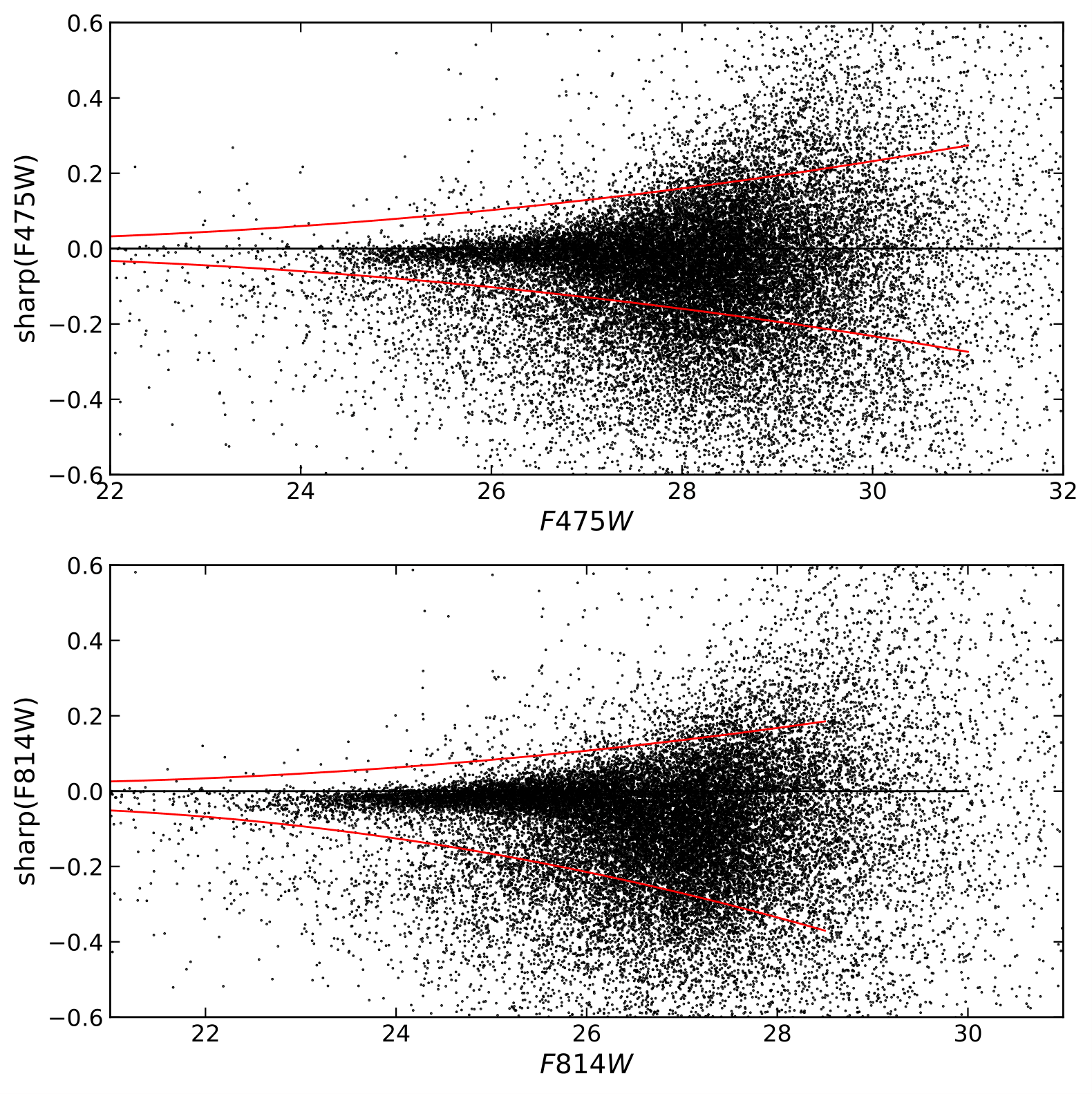}
    \includegraphics[width=.48\textwidth]{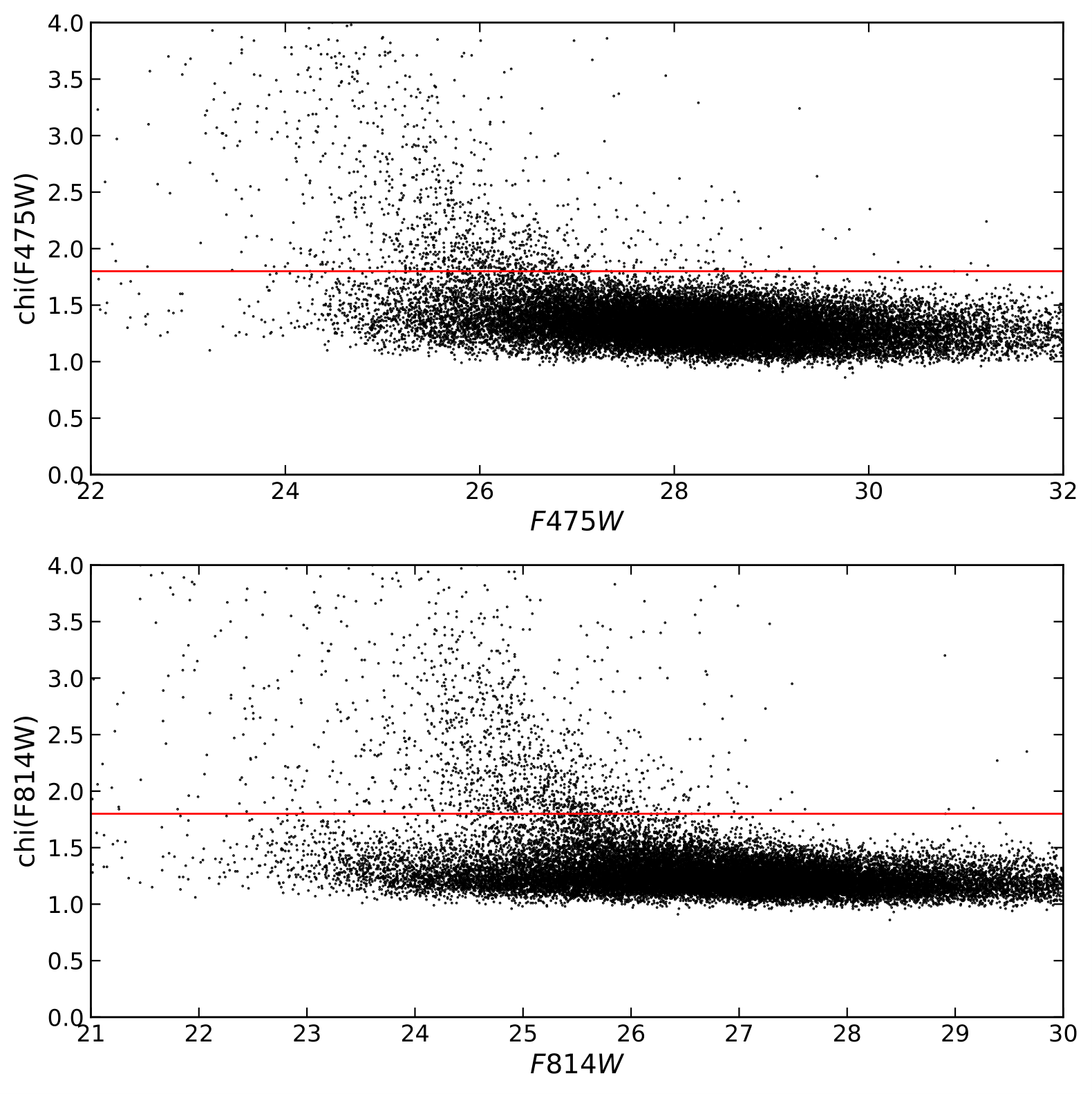}
    \caption{\emph{Left panel:}  \texttt{DOLPHOT} values of \texttt{sharp} plotted versus magnitude, for each of the two filters.  All detected objects are plotted here before any culling.  Non-stellar, extended objects scatter downward to lower \texttt{sharp} values.  The red lines denote the culling boundaries.  \emph{Right panel:}  \texttt{DOLPHOT} values of \texttt{chi} versus magnitude.  {Normally $\chi-$values near 1 indicate objects that ideally match the point-spread function, while higher $\chi-$values indicate objects that are poorly fit by the PSF.}}
    \label{fig:sharp_chi}
\end{figure*}

ASTs are the central tool for measuring completeness, and series of tests built on millions of artificial stars are now commonly found in the literature.  The default first-order approach to running ASTs is to generate an artificial-star population randomly and uniformly distributed in magnitude, color, or spatial location, and for many situations this approach may be quite adequate.  However, several studies place emphasis on generating distributions of artificial stars that mimic the luminosity function of the real objects in the image, or their distribution across the color-magnitude diagram  \citep[e.g.][]{stetson_harris1988,depoy+1993,bolte1994,aparicio_gallart1995,grillmair+1996,olsen+2003,harris+2006,hansen+2007,monachesi+2011,alamo-martinez+2013,williams+2013,harris+2016,cohen+2020}.  In other studies where the degree of crowding changes across the field, an analogous emphasis is put on matching the spatial distribution of the real objects \citep[e.g.][]{mateo1988,bolte1989,bolte1994,gallart+1996,radburn-smith+2011}.  

Aside from magnitude, which of the other features of an image are important depends very much on the particular situation.  To mention only a few of many possible examples, for stellar photometry studies in nearby galaxies \citep[e.g.][]{monelli+2010,williams+2014,cohen+2020}, much effort is concentrated on the effects of crowding. By contrast, in the ACS Virgo Cluster Survey of globular clusters (GCs) around the Virgo galaxies, \citet{jordan+2007} determine completeness fraction versus GC magnitude, sky background, and object morphology (the cluster scale size $r_h$), and then interpolate in the resulting 3D lookup table to find the recovery probability for any real GC in the list. \citet{adamo+2017}, for the LEGUS program on young star clusters in nearby galaxies, construct similar prescriptions for completeness versus object scale size.
\citet{peng+2011}, in their study of GC populations in the Coma cluster, derive the completeness at any spatial location in the survey by blanketing the area with artificial-star populations, implicitly accounting for magnitude, color, and background brightness. 
Cases involving severe crowding have generated a large previous literature by themselves \citep[e.g.][to mention only a  few]{stephens+2001,olsen+2003,olsen+2006,williams+2014,cohen+2020,schlafly+2019,saydjari+2023}.  In the extreme, photometry built on deconvolution methods has shown improvements in depth and precision over more normal PSF-fitting photometric codes \citep[e.g.][]{lauer+2012,dong+2018}. 

In still other studies, a significant concern is the presence of a strong gradient of local sky brightness across the field (quite independently of crowding), for example in the measurement of star cluster populations around giant galaxies \citep[see, e.g.][for recent examples]{whitmore+2010,faifer+2011,escudero+2015,escudero+2018,caso+2017,caso+2019,ennis+2020}. Here, the issue has traditionally been dealt with by deriving completeness curves versus magnitude in distinct radial zones centered on the galaxy.  This approach implicitly but roughly accounts for the radial gradient of sky background noise, which decreases with increasing galactocentric distance.  For photometry within star clusters or in nearby galaxies, completeness can be a highly localized function of location, since areas near very bright stars are strongly compromised \citep[e.g.][]{bedin+2008,williams+2014}.

In almost all of these previous studies, the completeness is estimated from ASTs binned in magnitude. The resulting trend may either be numerically interpolated across bins, or else fitted with some smooth analytic function.  Examples of such functions, all of which have basically similar shapes, can be found in \citet{fleming+1995}, \citet{puzia+1999}, \citet{barker+2004}, \citet{alamo-martinez+2013}, \citet{harris+2016}, and \citet{harris_reinacampos2023}.

For any binned data, however, bin-jumping may need careful attention.  
\citet{alamo-martinez+2013} demonstrate especially clearly that in cases where the real objects follow a luminosity function (LF) that rises steeply toward fainter magnitudes, and where the measurement uncertainty is also inevitably rising with fainter magnitude, asymmetric bin-jumping can occur (Eddington bias) along the downturn of the completeness curve.  \citet{saydjari+2023} also discuss this effect, which can even produce the interesting anomaly of a completeness level $f$ larger than 1.0 in bins just brighter than the downturn region.  As these discussions emphasize, this effect places importance on matching the LF of the inserted artificial stars to the real stars, in order to get correct relative numbers of bin-jumping stars at each magnitude level.  

As a basis for developing a new model for photometric completeness, we draw from two different sets of optical/NIR photometry from the Hubble Space Telescope (HST) that are deliberately chosen to bring into play all the parameters described above.  These are laid out in the next sections.

\section{Test Case 1: NGC 1275 and its Globular Clusters} \label{ngc1275}

NGC 1275 is the giant and extremely active early-type galaxy (ETG) at the center of the Perseus cluster \citep{harris+2020}.  As is the case for all giant ETGs, NGC 1275 has a rich population of many thousands of classic old GCs spread throughout its halo \citep{carlson+1998,penny+2012,harris_mulholland2017,harris2023}.  But unlike most such systems, its inner $\sim20$ kpc also has an extensive, massive network of gas and dust filaments that hosts a large population of young stars and star clusters \citep{canning+2010,canning+2014,lim+2020}.  At the Perseus distance of 75 Mpc the star clusters are unresolved and appear near-starlike in structure even with the spatial resolution of HST \citep{harris+2020}.  This last feature fortunately allows stellar photometry codes like \texttt{DAOPHOT} \citep{stetson1987} or \texttt{DOLPHOT} \citep{dolphin2016} to be used very effectively (cf. the references cited above).

\begin{figure*}
    \centering
    \includegraphics[width=.85\textwidth]{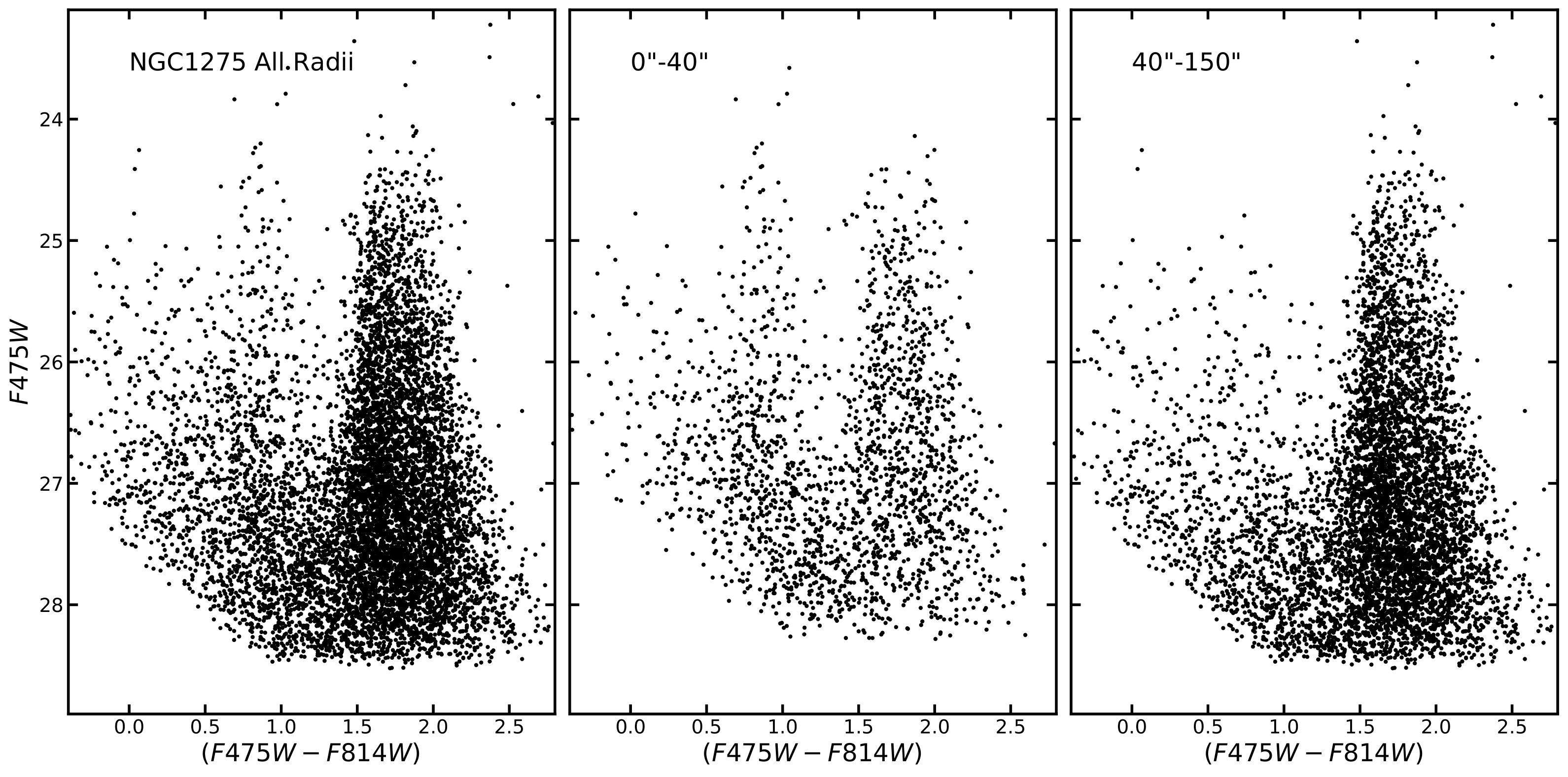}
    \caption{Color-magnitude diagram (CMD) for the star cluster population around NGC 1275.  No correction has been made for foreground reddening.  The relatively sharp cutoff of the data at the faint end represents the signal-to-noise ratio ${\rm SNR} = 4$.  {The second and third panels show the CMD for inner and outer radial zones:  in the inner zone, the bluer clusters with (F485W-F814W) $\lesssim 1.2$ are relatively more prominent, and the limiting magnitude of the photometry is brighter.}}
    \label{fig:cmd}
\end{figure*}

\subsection{HST Photometry}

The photometry of the NGC 1275 cluster system is taken from the raw data described in \citet{harris+2020} and the field is shown in Figure \ref{fig:ngc1275}.  These HST images include exposures in the filters (F475W, F814W) from the ACS/WFC camera, one orbit per filter and three sub-exposures per orbit, with total exposure times of 2436 sec in F475W and 2325 sec in F814W.  Previous photometry of the system is reported in \citet{harris2023} but the  measurements were redone for the present study through the photometric packages in \texttt{DOLPHOT}.  The reference image used to detect objects for the photometry was adopted as the deep \texttt{*.drc} F814W image from the MAST Archive, where the three individual exposures were combined through \texttt{AstroDrizzle}.  

\texttt{DOLPHOT} is a versatile code especially tuned for HST imaging data, and detailed discussions of its many input parameters can be found in, e.g., \citet{dalcanton+2009}, \citet{monelli+2010}, \citet{radburn-smith+2011}, \citet{williams+2014}, and \citet{cohen+2020}.  A number of the parameters are important particularly for photometry in extremely crowded fields.  However, as will be seen below, the star cluster population in and around NGC 1275 is not strongly crowded across most of the field and so the choices are less critical.\footnote{The basic parameters adopted here were $\texttt{RAper} = 6.0$\,px with centroiding; $\texttt{RSky} = (15, 25)$\,px for the sky annulus; $\texttt{FitSky} = 2$; $\texttt{SigFind} = 3.0$; and a Lorentz profile for the point spread function (PSF).  The results were insensitive to these choices.}  Objects were detected on the original reference image, without removal of the overall light profile of the galaxy.

From the raw \texttt{DOLPHOT} output measurements, any objects not detected on both filters were rejected, as were objects with signal-to-noise ratio ${\rm SNR} < 4$, or ones that were non-starlike in morphology.  The latter culling step was done with the \texttt{DOLPHOT} \texttt{chi} and \texttt{sharp} parameters as illustrated in Figure \ref{fig:sharp_chi}.  These figures show the distribution of detected objects before any culling is done.  Objects lying outside the boundary lines shown in the figure are mostly faint background galaxies that are visibly nonstellar, extended or asymmetric, and thus with excessively negative \texttt{sharp} values; these were rejected.   At fainter magnitudes the \texttt{sharp} values for stars inevitably spread more, so the rejection boundaries were drawn to reflect the observed scatter.  The boundary lines were set fairly conservatively because many of the GCs are very slightly more extended than the pure PSF and thus have small negative \texttt{sharp} values and $\chi >1$; see \citet{harris+2020} for a detailed discussion and measurement of cluster sizes. The brighter objects in the final culled list of objects are marked in Fig.~\ref{fig:ngc1275}, and their distribution in the color-magnitude diagram (CMD) is shown in Figure \ref{fig:cmd}.  

\begin{figure}
    \centering
    \includegraphics[width=.48\textwidth]{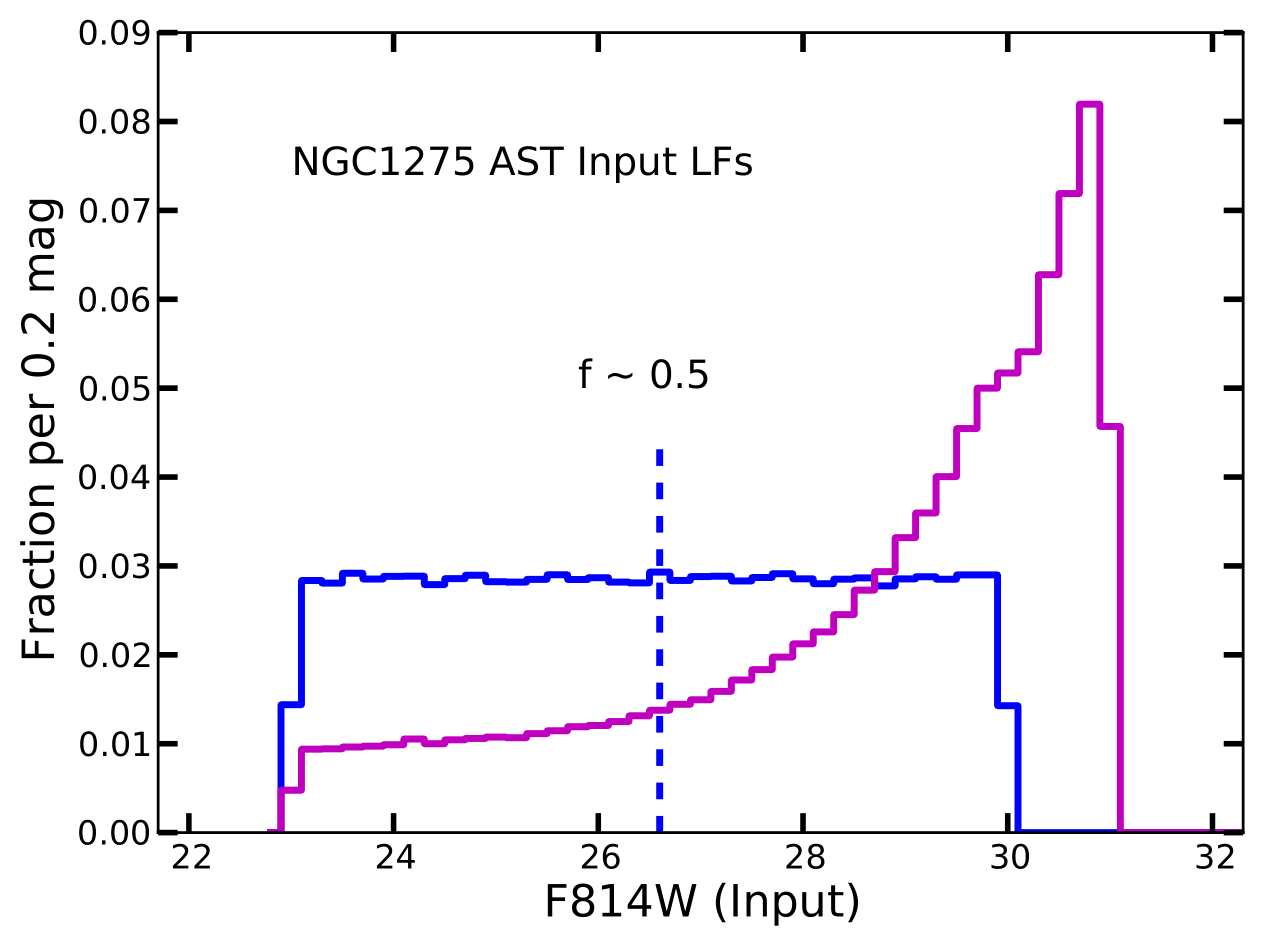}
    \caption{Luminosity functions (LFs) for the two sets of artificial-star tests run on the NGC 1275 images.  The first set (blue) assumes a flat LF while the second set (magenta) assumes an exponentially rising LF towards fainter magnitudes.  The vertical dashed line indicates the magnitude level at which the recovery fraction is approximately 50\% depending on radial zone and color (see text for additional details).}
    \label{fig:inputlf}
\end{figure}

 {Classic old GCs are found in large numbers in all giant galaxies studied to date \citep[e.g.][]{larsen+2001,peng+2006,harris2009,harris+2016,harris+2017,harris2023,hartman+2023}.  Their range of colors is primarily driven by GC metallicity, and the color distribution function (CDF) usually can be seen to have distinguishable `blue' and `red' vertical sequences in the CMD that correspond to mean metallicity values near [Fe/H]  $\simeq -1.5$ (blue) and $-0.5$ (red).  In Fig.~\ref{fig:cmd} these sequences are centered near the colors 
 ${\rm F475W} - {\rm F814W} \simeq 1.6$ and 2.0. In giant galaxies particularly, these two subgroups each have broad enough internal ranges of metallicity that the sequences overlap in color  \citep[see][for numerous examples and statistical tests]{larsen+2001,peng+2006,harris2023,hartman+2023}.} However, for NGC 1275 an additional feature not seen in most other giant ETGs is the presence of many young blue clusters (color indices $\simeq 0.6 - 1.2$ in Fig.~\ref{fig:cmd}), which are found in the inner halo and are associated with the gas and dust filaments as mentioned above.  These clusters increase the total range of colors in the entire sample of objects, which is a useful feature of the data for our testing purposes.
 
 \subsection{Artificial-Star Tests}
 
To provide the raw data for evaluating completeness, two sets of artificial stars following different input luminosity functions (LFs) were used.  The first set consisted of $2.6 \times 10^5$ artificial stars that randomly and uniformly covered the magnitude and color ranges $23 < {\rm F814W} < 30$, $0.0 < {\rm F475W}-{\rm F814W} < 3.0$ (i.e. a flat LF).  The second set consisted of $2.0 \times 10^5$ stars following an exponentially rising LF $n(m) \sim e^{a m}$.  To generate the latter over an adopted magnitude interval $(m_0, m_1)$, first $n$ numbers \{$q_i$\} were produced following a random uniform distribution from 0 to 1.  Then the magnitudes \{$m_i$\} given by the transformation
\begin{equation}
    m_i = m_0 + {1 \over a} \, {\rm ln} \left[1 + q_i \left(e^{a(m_1 - m_0)} - 1\right) \right]
    \label{eq:lf}
\end{equation}
are distributed exponentially as desired.  For this case we use $a=0.7$, representing quite a steep rise.
 
Both input LFs are shown in Figure \ref{fig:inputlf}.  In each case the fake stars were inserted one-by-one into the images, and then remeasured through \texttt{DOLPHOT} with exactly the same culling criteria as described above. The fake stars were also spatially distributed following a centrally concentrated radial profile that generated roughly the same numbers of stars within radial annuli of the same width.  

\begin{figure}
    \centering
    \includegraphics[width=.48\textwidth]{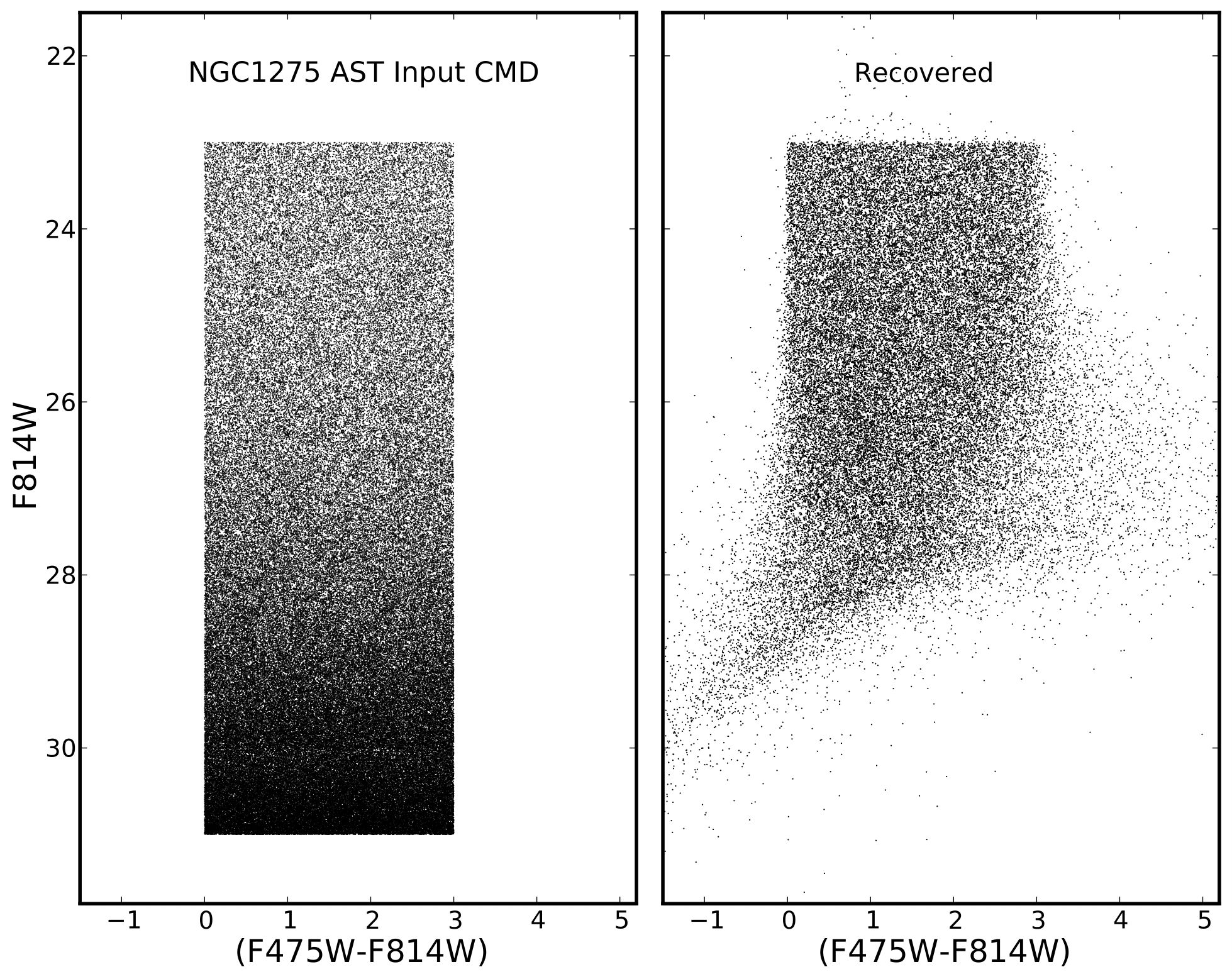}
    \caption{Color-magnitude diagram in F814W versus F475W$-$F814W for artificial stars following the exponentially rising LF.  \emph{Left panel:} The input magnitudes and colors of the artificial stars.  \emph{Right panel:} Measured values for stars that were successfully recovered (i.e. were detected and passed the culling tests).}
    \label{fig:fakecmd}
\end{figure}
 
In Section \ref{background}, emphasis was placed on the need to design ASTs that should match the LF of the target objects. The two input LFs were chosen to roughly bracket the expected LF of the real data.  For the NGC 1275 GC population, the LF  should not necessarily follow rising numbers towards fainter magnitude, because the real data have a completeness limit near the classic turnover of the GC LF at ${\rm F814W} \sim 27$ \citep{harris+2020}, so at fainter magnitudes the numbers of GCs will actually \emph{decrease} past the completeness limit.  However, this decline will be offset by the numbers of faint background contaminants, which rise at fainter magnitudes.  
 
In Figure \ref{fig:fakecmd}, the CMDs of both the input and recovered measurements are shown for the rising LF.  Near a magnitude of ${\rm F814W} \simeq 28$, a transition zone is apparent through which the numbers of objects that were successfully recovered gradually die away either because they are too faint to be detected in one filter or the other, or else detected but then rejected on the basis of low SNR, high \texttt{chi}, or high \texttt{sharp} values.

\begin{figure}
    \centering
    \includegraphics[width=.46\textwidth]{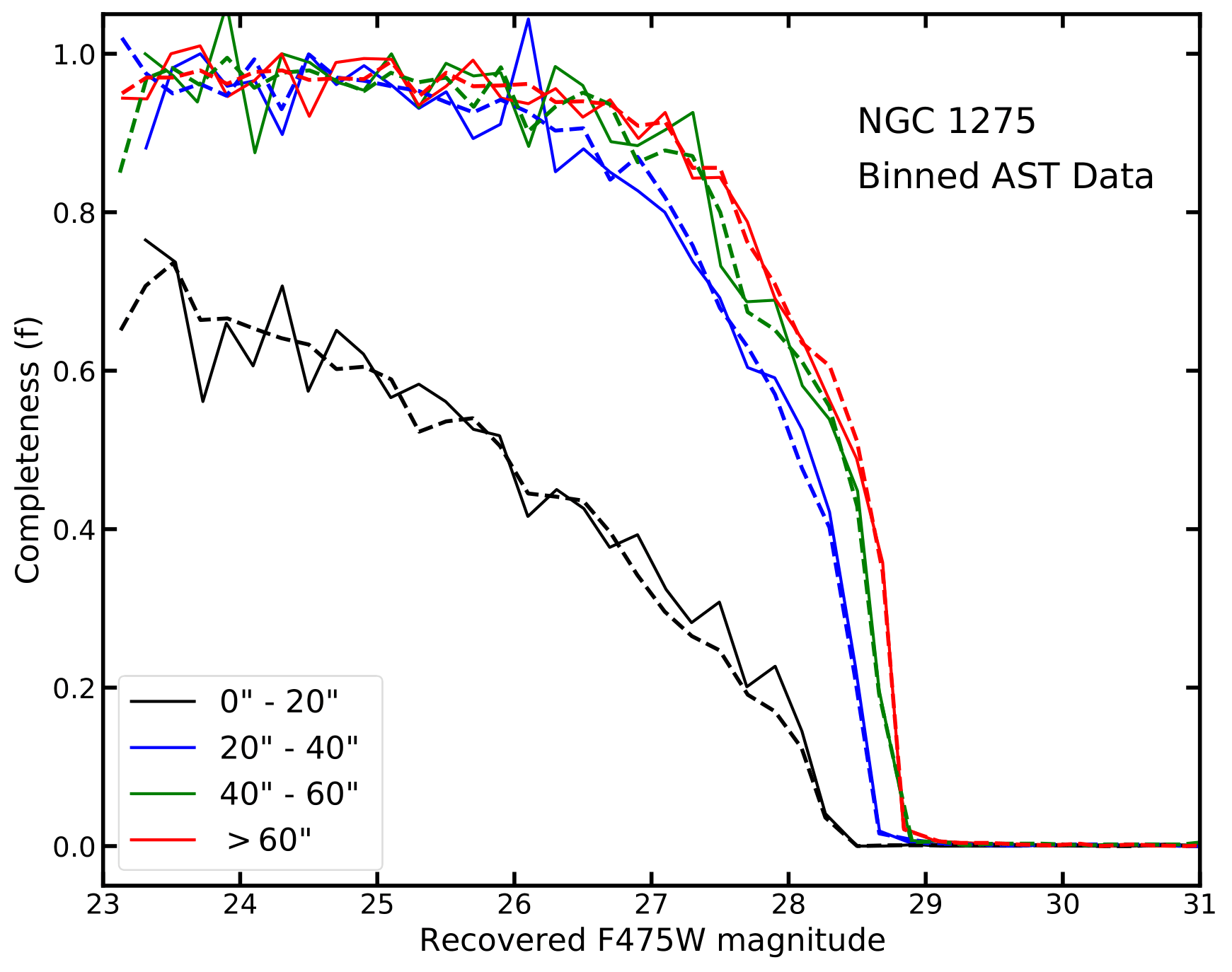}
    \caption{Completeness as a function of magnitude for the artificial stars in four radial zones around NGC 1275, for the AST data binned in 0.2-mag steps. \emph{Solid lines} show the results for the exponentially rising input LF, while the \emph{dashed lines} show the results for the flat input LF.  Black lines are for radial zone 1 ($r < 20''$), blue for zone 2 ($20''-40''$), green for zone 3 ($40''-60''$), and red for zone 4 ($r > 60''$). Zone 1 is heavily affected by background sky noise in the central bulge of the galaxy, reducing the completeness fraction even for relatively bright objects.  {Note that at brighter magnitudes, the curves for the rising LF are statistically noisier because the numbers of input stars are smaller than for the flat LF.}}   \label{fig:fcurve}\end{figure}
 
 The AST data were binned in steps of 0.2 mag, and the completeness fraction $f(m)$ in each bin was plotted versus magnitude.  We illustrate the results using the ${\rm F475W}$ magnitude (though either magnitude could be used equally well). The curves for four different radial zones around the center of the galaxy, and for both input LFs, are shown in Figure \ref{fig:fcurve}.  For the innermost zone ($r < 20'' = 400$ px), even moderately bright stars are incompletely recovered because of the very bright and complex background light from the central bulge of the galaxy.  Outside this zone, however, the background surface brightness of the galaxy falls away rapidly, and the completeness curves converge more closely together.   As seen in the Figure, the completeness curve results are similar for both input LFs.
 The stronger effect is the change from one radial zone to the next, pointing to the desirability of a general solution depending on local sky brightness or sky noise that will not impose radial binning. 

\section{Test Case 2:  NGC 3377 and its Halo Stars} \label{ngc3377}

\begin{figure}
    \centering
    \includegraphics[width=.46\textwidth]{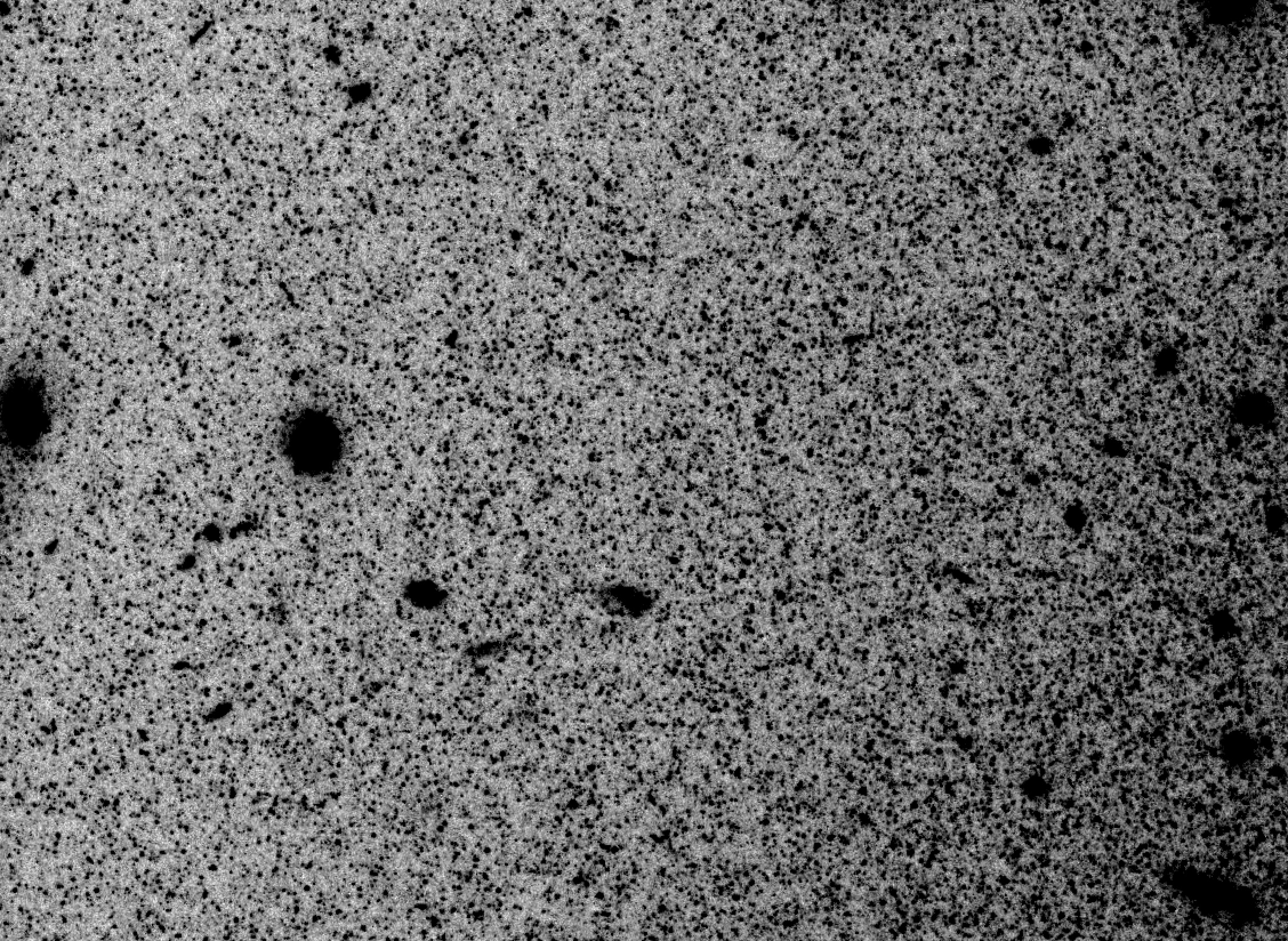}
    \caption{A portion of the field in the halo of NGC 3377, imaged with HST/ACS.  The field shown is $55''$ across  (about one-quarter of the width of the ACS field of view), corresponding to a linear distance of 3 kpc. Large numbers of halo red-giant stars are scattered across the image; the galaxy center is off the field to the lower right.  See also Figures 1 and 2 of \citet{harris+2007}.}
    \label{fig:n3377field}
\end{figure}

The NGC 1275 data have suitably wide ranges in color and local sky noise, but they do not enable strong tests of the crowding parameter.   A useful example of a single field that displays a larger range of crowding is the HST ACS deep imaging of the halo of NGC 3377.  This galaxy is an intermediate-sized ETG in the Leo group at $d=11$ Mpc, at high Galactic latitude ($b=58^o$) and low foreground extinction ($A_V = 0.09$).  The raw HST images comprise total exposures of 38500 sec in F606W and 22260 sec in F814W and were originally used to measure the metallicity distribution of the resolved red-giant-branch (RGB) stars in the halo \citep{harris+2007}.  Since NGC 3377 is not a giant galaxy, its scale size is small enough ($R_{eff} \sim 2$ kpc $\simeq 38''$) that the degree of crowding changes strongly across the ACS field of view moving outward from the galaxy center.  A 3-kpc portion of the field is shown in Figure \ref{fig:n3377field} where the gradient of surface brightness is clearly visible \citep[see also Figures 1 and 2 of][]{harris+2007}.

\begin{figure*}
    \centering
    \includegraphics[width=.9\textwidth]{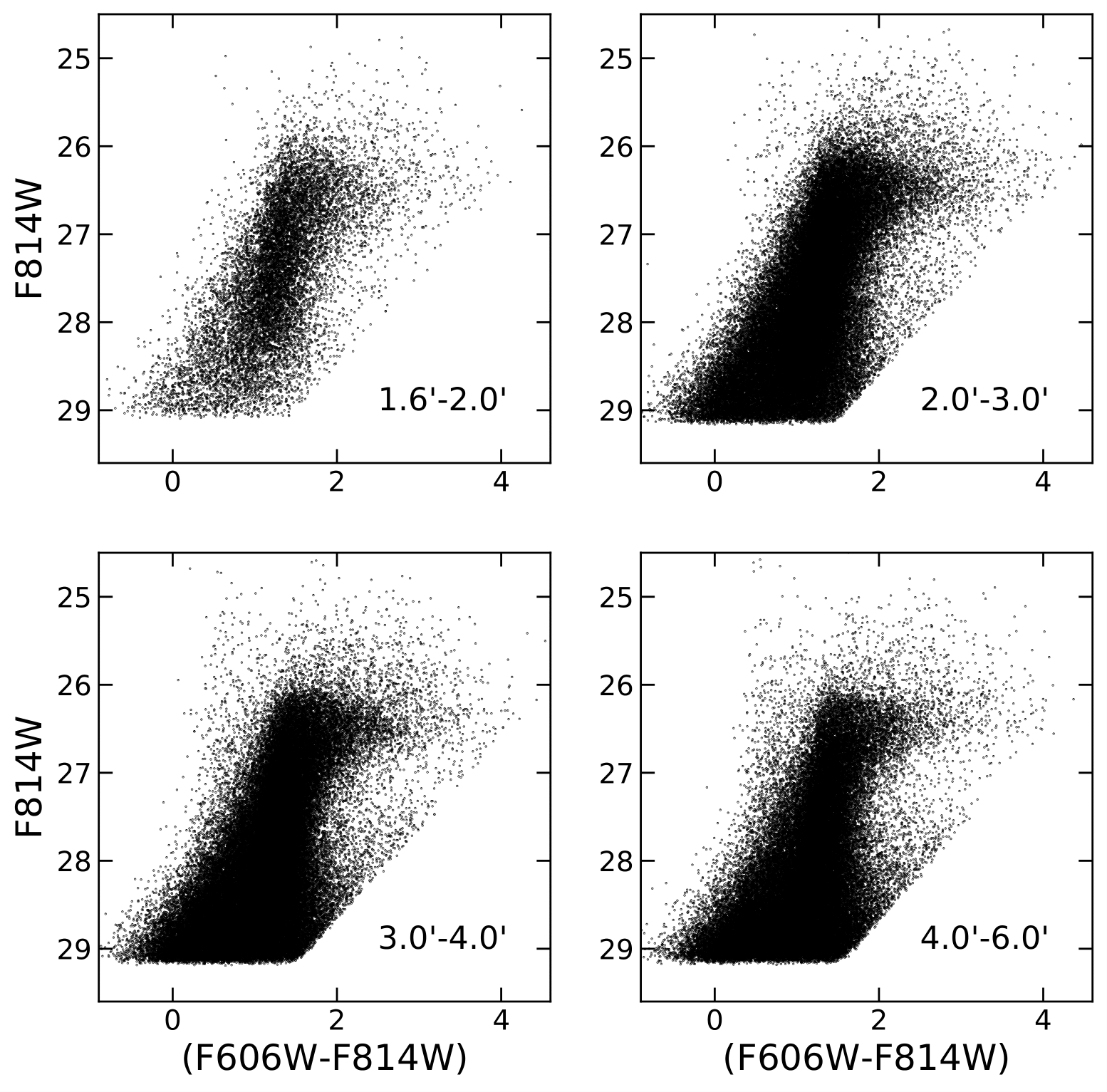}
    \caption{Color-magnitude diagrams for the NGC 3377 halo, in four different zones of projected galactocentric radius $R_{gc}$.  The innermost edge of the ACS field is at $1.6'$ from the galaxy center.}
    \label{fig:cmd_4panel}
\end{figure*}

Our new \texttt{DOLPHOT} reductions for the NGC 3377 field are shown in Figure \ref{fig:cmd_4panel}. At all galactocentric radii the CMD of the recovered objects is completely dominated by many thousands of halo RGB (red-giant-branch) stars with negligible field contamination.  In all radial zones, the sharp lower boundary of the CMD is due to the SNR limit imposed in the selection stage.  

To set up the data for the subsequent modelling, a set of $5 \times 10^5$ artificial stars distributed uniformly across the field was used, covering the magnitude range ${\rm F475W} = 23$ to $36$ and color range ${\rm F606W}-{\rm F814W} = -1$ to $5$, and following an exponentially rising luminosity function with $a=0.7$ (Eq.\ (\ref{eq:lf})) to match a normal old RGB LF.

An important difference between the NGC 3377 and NGC 1275 datasets is that for NGC 3377, the RGB star population itself is the dominant source of the local sky noise. Separate parameters for sky noise and crowding are thus not needed and we use only the crowding parameter to represent both.  The specific way that crowding is defined is described in the next section.

\section{Model Parameters} \label{model_params}

Recasting the completeness problem in terms of recovery probability $p$ in a (generalized) linear model has two major advantages:  any issues connected with binning of the data are bypassed, and as will be seen below, all the physical variables can be treated \emph{simultaneously} within a single solution. 
The next step is to define more specifically the variables \{$x_j$\} determining recovery probability that will actually be used. We will consider the following potential variables:
\begin{enumerate}
    \item The first variable $x_1$, and usually the most important one, is \emph{magnitude}.  For the two sample databases described above, in principle the magnitude in either of the two filters defining the CMD could be used equally  well.  For NGC 1275 we use the F475W magnitude, while for NGC 3377 we use F814W.
    \item The second variable is the \emph{color index}, in this case $x_2 = {\rm (F475W}-{\rm F814W)}$ for NGC 1275 or $x_2 = {\rm (F606W}-\rm{F814W})$ for NGC 3377.
    \item The third variable $x_3$ represents local sky brightness or \emph{background noise}, which in turn determines the detection threshold.  This parameter can be quantified in several ways and is discussed further below.  
    \item The fourth variable $x_4$ represents \emph{crowding}.  This parameter can also be defined in different ways and is discussed further below.  
    \item There is a potential fifth variable $x_5$, which represents object \emph{morphology} or \emph{scale size}; essentially, objects that are more extended or of lower surface brightness are harder to detect than point sources at the same magnitude and color.  This might be quantified in a simple way by (e.g.) the half-light radius or mean surface brightness of an object \citep[e.g.][]{jordan+2007}.  In the present study, however, we specifically measure only starlike, unresolved objects, so $x_5$ is not used.  It is worth noting as well that we are working with HST imaging data where the stellar PSF is {closely similar for all the images for a given filter in the same observing program, unless extremely different epochs are involved (not the case here)}.  For ground-based imaging, the PSF will often differ between images, and thus even for stellar photometry the parameter $x_5$ will need to be introduced.  In this case it could be quantified as the FWHM of the PSF.
\end{enumerate}

\begin{figure}
    \centering
    \includegraphics[width=.48\textwidth]{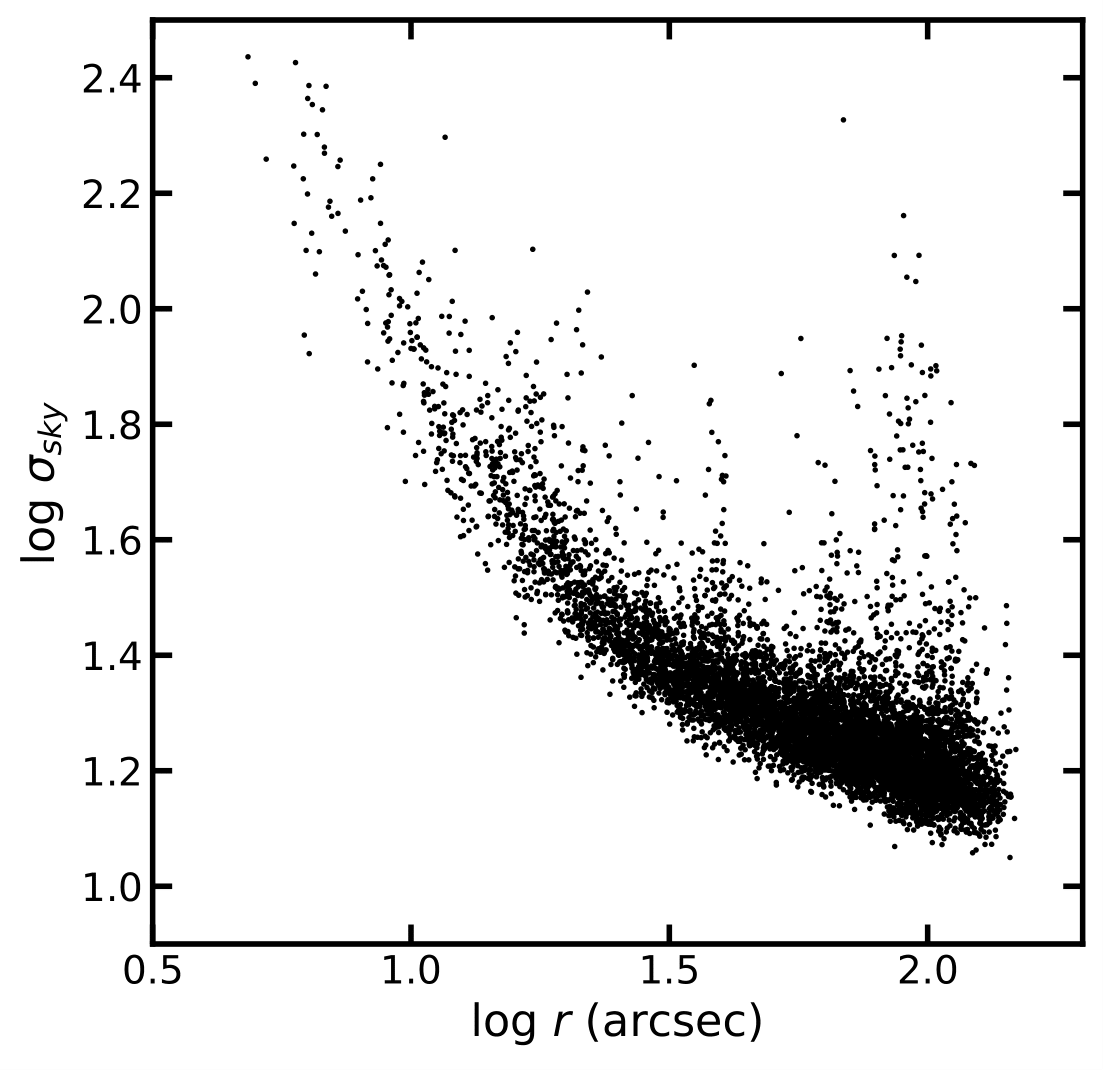}
    \caption{The standard deviation of local sky noise plotted versus galactocentric radius, for the measured objects around NGC 1275.  Here $\sigma_{sky}$ in adu is measured on the F814W reference image within an annulus of 4 to 9 px around each object.  Note the `spikes' of higher noise at large radius where satellite galaxies are located.}
    \label{fig:sigrad}
\end{figure}

\subsection{Background noise}  

The parameter $x_3$ requires some additional discussion.  For the NGC 1275 field and in many previous studies of this type (cf. the references cited earlier), the radius $r$ from the center of the galaxy may be used essentially as a proxy for local sky brightness and thus for the noise $\sigma_{\rm sky}$.  These quantities may be well correlated (see Figure \ref{fig:sigrad}), though not always by a simple scaling.  More importantly, defining $x_3$ in terms of $r$ does not account for morphological asymmetry or irregularities in the central galaxy, or for large satellite galaxies in the field (see Fig.\ \ref{fig:ngc1275}) that generate their own local enhancements of surface brightness.  Extreme cases where $\sigma_{\rm sky}$ differs strongly with location but in a complex way, such as photometry of stars on a nebular background or within a star cluster \citep[cf.][]{bedin+2008}, should also be manageable in a general procedure.  Defining $x_3$ in terms of local sky noise goes directly to the essential quantity that determines the detection threshold, and bypasses any of these geometric or location-specific limitations.  

Fortunately, the relevant measurements can easily be done with standard photometry codes.  
For the NGC 1275 field, after some experimental trials, we adopted  
$x_3 ={\rm ln}\ \sigma_{\rm sky}$, where $\sigma_{\rm sky}$ is the standard deviation of the sky noise immediately around each object.\footnote{{Note that Fig.~\ref{fig:sigrad} plots log$_{10}$ $\sigma_{\rm sky}$, though for numerical convenience we use ${\rm ln}\,\sigma$ in the LR solution.}}  Specifically, the deep F814W reference image was used to directly measure the local value of $\sigma_{\rm sky}$ within an annulus of 4 to 9 px (2.0 to 4.5 times the stellar FWHM) at the location of each object.  To minimize any effects of crowding, the value of $\sigma_{\rm sky}$ was taken to be the mode of the rms scatter of pixel values within the annulus after up to 10 iterations of $3\sigma$ clipping.  

\begin{figure}
    \centering
    \includegraphics[width=.48\textwidth]{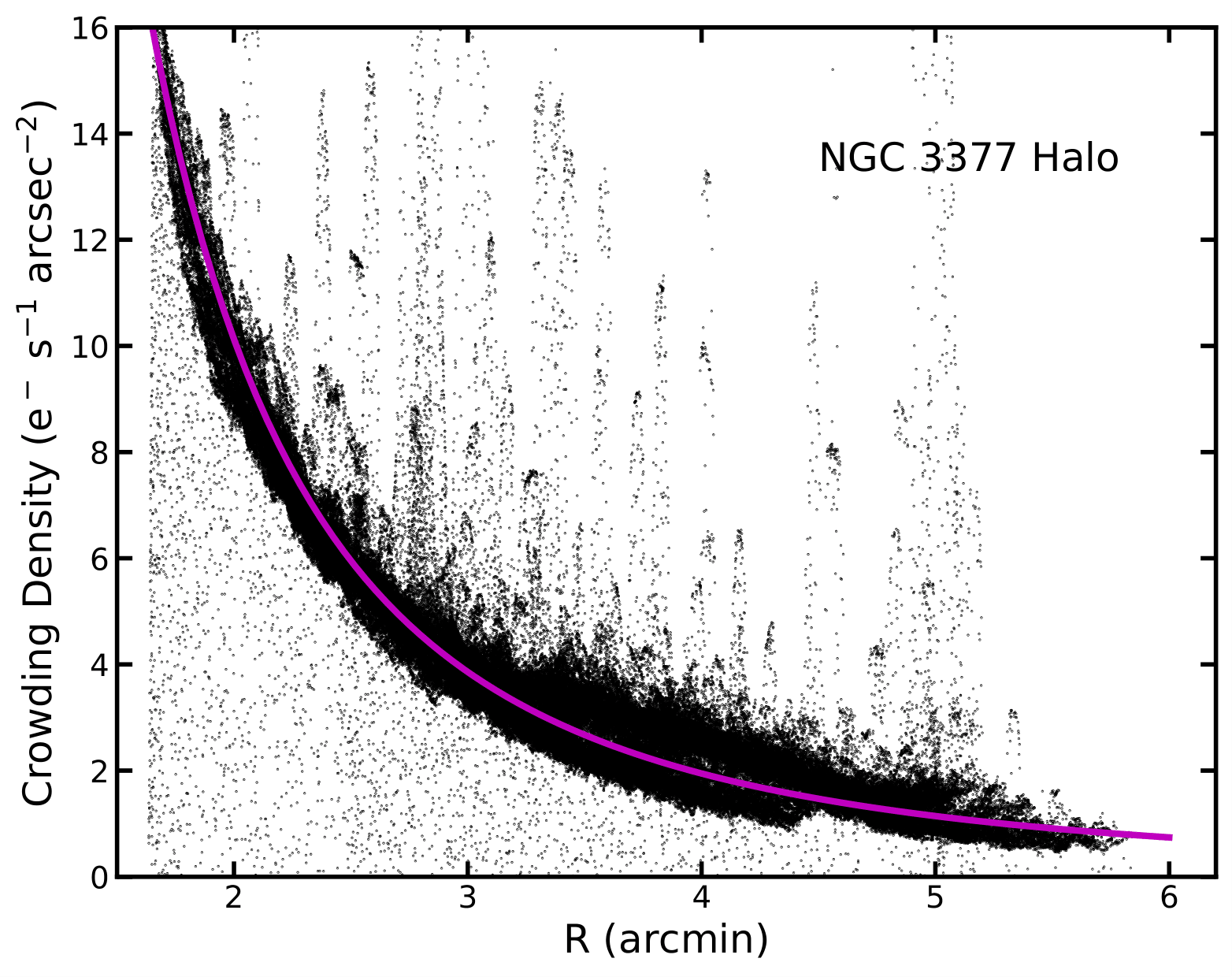}
    \caption{Dependence of local crowding level on galactocentric radius $R_{gc}$ (arcmin), for all measured stars in the NGC 3377 field.  The crowding density parameter $x_4$ is defined as the local mean surface brightness in the image around each object after background subtraction, as described in the text.  The trend follows a simple power-law form $x_4 \sim R^{-2.38}$ shown by the magenta line, and is equivalent to the surface brightness profile of the galaxy.  {The `spikes' of points extending upward from the main locus represent objects near small satellite or background galaxies, or bright stars, where the local background light is enhanced.  The sparse dusting of points below the main locus shows objects near the CCD detector edges,}}
    \label{fig:crowding}
\end{figure}

\subsection{Crowding}

An intuitively clear definition of crowding is simply the local number density of objects brighter than a given magnitude \citep{williams+2014}.  Specifically, for the NGC 1275 field the number of real objects with F475W $\leq 27.0$ and within $r < 60$ px $= 3''$ of each object is counted (including the central object itself), and their density expressed as number per resolution element.  We take one resolution element to be a circle of radius equal to the PSF hwhm or $0.05'' = 1$ px, so the 60-px circle contains 3600 resolution elements. It should be noted that any star can be crowded not just by other stars but also by any other type of object such as small faint background galaxies, so all detected objects within the 60-px circle brighter than the adopted threshold are counted.

\citet{king+1968} or \citet{hogg2001} describe crowding in terms of the fractional area covered by nearby objects, i.e.\ the inverse of their number density.  A problem with either definition is that they will be affected by the photometric completeness itself, which is the very quantity we are trying to measure.  Also, at very high crowding levels the number density will enter a nonlinear regime where objects overlap each other. 

For the NGC 3377 field where local crowding can reach more extreme levels, an alternate definition is adopted:  we define $x_4$ in terms of the total measured light within a specified annulus around each star.  This definition implicitly includes all types of objects and is immune to incompleteness in number counts and overlapping stars.  It assumes only that the LF of the RGB stars will be consistent from place to place in the NGC 3377 halo.  

In Figure \ref{fig:crowding}, the crowding $x_4$ as defined this way is plotted versus projected galactocentric radius $R_{gc}$.  Specifically, $x_4$ equals the light in an annulus from 4 to 40 px ($0.2''$ to $2''$) around each star, as measured by the \emph{phot} module in \texttt{DAOPHOT}, directly from the F814W reference image.  For convenience $x_4$ is expressed in units of surface brightness, $x_4 = ({\rm SB}_{\rm F814W} - {\rm SB}_0)$\,$e^-\, s^{-1}$\,arcsec$^{-2}$ where $SB_0$ is the background sky brightness as measured in the uncrowded corners of the image at large radius.

{In brief, different ways to define a crowding variable are available but they may not all be equally effective.  The best choice will be not just one easily measurable but one most suited to the needs of the individual situation.}

\section{Defining the Model} \label{lrmodel}

With the testbed data in hand, the next stage is to develop a framework for modelling the recovery probability $p$ as a function of the variables listed above.  Based on our earlier definition, any star is said to be \emph{recovered} by the photometry if it is first successfully \emph{detected} (it lies above the detection threshold) but also then successfully \textit{passes} any following culling tests like those described above (detected in both filters, more than the minimum SNR, starlike morphology, etc.).

\begin{figure}
    \centering
    \includegraphics[width=.48\textwidth]{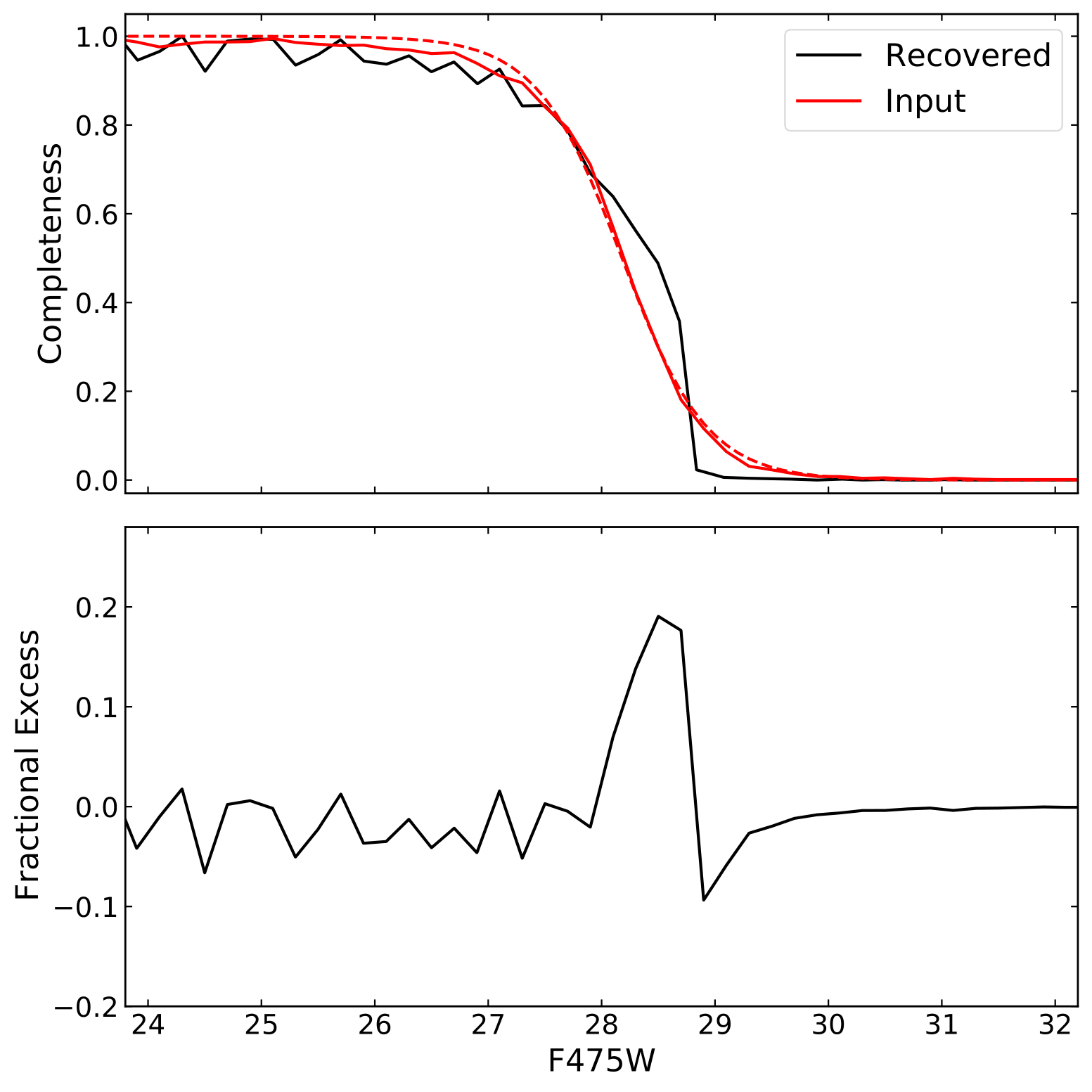}
    \caption{\emph{Upper panel:} Comparison of completeness fraction plotted versus either the measured magnitudes (black line) or the input magnitudes (red line).  See text for the exact definitions of each fraction.  The dashed red line shows the best-fit sigmoid function fit to the binned $f$ versus input magnitude.    \emph{Lower panel:}  Fractional excess of recovered stars for each bin, gained at the expense of other bins:  see text for explanation.}
    \label{fig:fcompare}
\end{figure}

\subsection{Logistic Regression} \label{lr}

Though it will not be our final version, the LR model in its basic form provides a good starting point.  If we have a vector of variables $\mathbf{X} = \{ x_i \} $ that correspond to the physical properties listed above\footnote{We will assume the convention that there is a constant value of $x_0 = 1$ for all observations in order to avoid listing a constant offset in all subsequent equations.}, we can model the probability of recovery as a generalized linear model 
\begin{equation}
    y = g(p) = \ln\left(\frac{p}{1-p}\right) = \bm{\beta}\mathbf{X}
    \label{eq:logit}
\end{equation}
where $g(p)$ is the \textit{logit} function and $\bm{\beta} = \{ \beta_i \}$ are a set of corresponding coefficients that must be solved for (and give the relative importance of each factor). The LR expression for the probability of recovery is then \citep{eadie+2022}
\begin{equation}
    p = g^{-1}(\bm{\beta}\mathbf{X}) = \frac{1}{1+e^{-\bm{\beta}\mathbf{X}}}
    \label{eq:invlogit}
\end{equation}
where $g^{-1}(x)$ is the \textit{logistic} function (i.e. the inverse of the logit).

The simplest possible case would be stellar photometry in a single filter, in an uncrowded field, over a uniform sky background.  In that case the recovery probability reduces to 
\begin{equation}
    p = \frac{1}{1+e^{-({\beta_0 + \beta_1 x_1)}}} ,
      \label{eq:simplelr}
\end{equation}
leaving only the zero-point $\beta_0$ and magnitude coefficient $\beta_1$ to be solved for.  As noted previously, this simple sigmoid-curve version does provide an approximate match to observed $f(m)$ completeness curves in situations where the other factors are not changing, as in Eq.\ (\ref{eq:fcurve}) above.  Setting $\alpha = -\beta_1$, $m_0 = -(\beta_0/\beta_1)$ makes Eqs.\ (\ref{eq:fcurve}) and (\ref{eq:simplelr}) equivalent.

 \begin{figure*}
    \centering
    \includegraphics[width=0.96\textwidth]{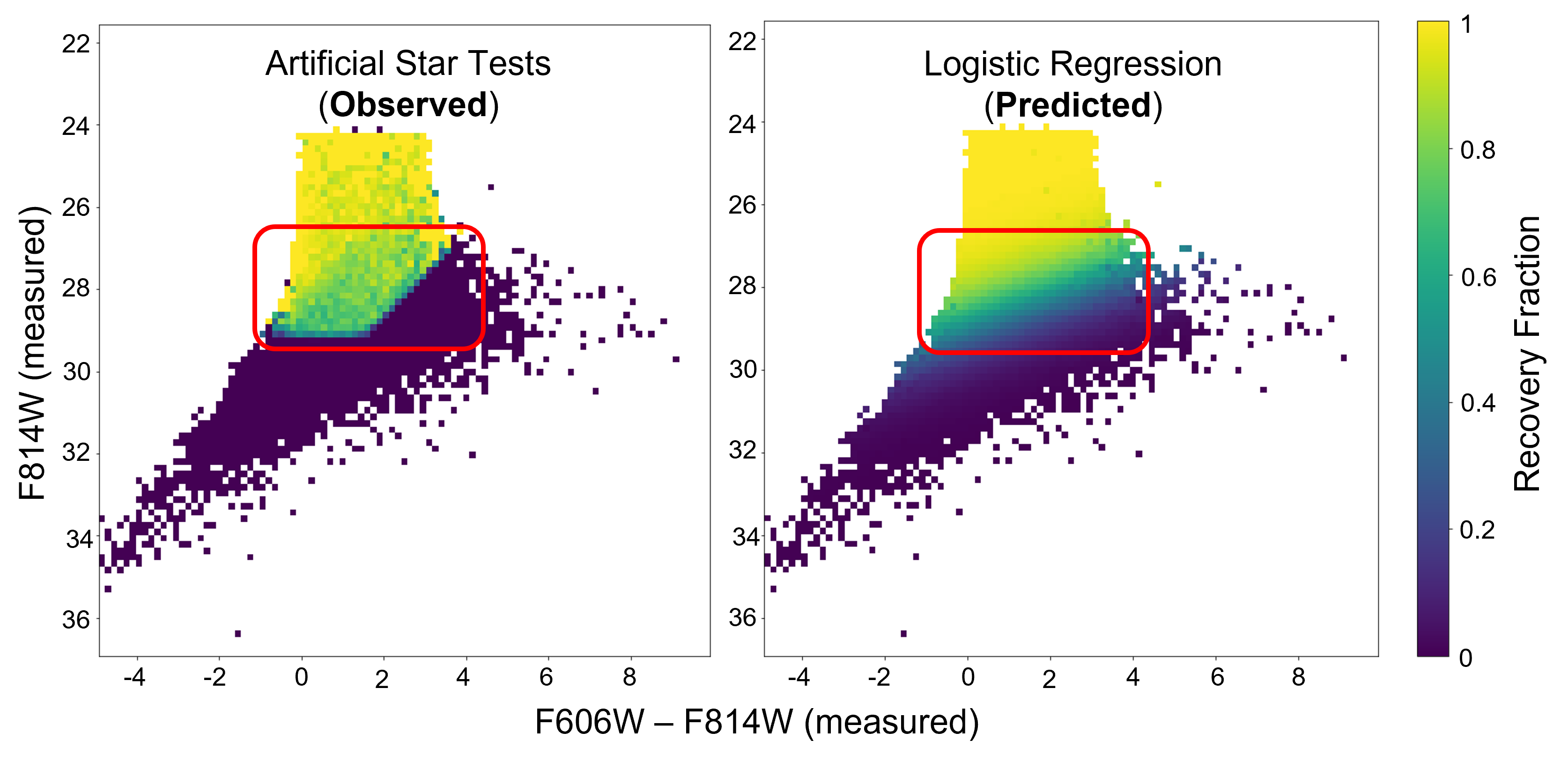}
    \caption{Demonstration of the inadequacy of a simple LR model for NGC 3377. The left panel shows the average recovery fraction binned across the CMD, which highlights sharp cutoffs in colour and magnitude as well as asymmetric complex behaviour (highlight with a red box). The right panel shows the predictions from a simple LR model which incorrectly smooths over both of these behaviours, leading to an overly broad completeness transition that is ``tilted'' in the CMD.}
    \label{fig:lrcompare}
\end{figure*}

The simple LR model in Eq.\ (\ref{eq:invlogit}) can be used as it stands with the complete set of variables \{$x_i$\} to obtain approximate solutions that will give the recovery probability $p_i$ for each real object $i$ as a function of its magnitude, color, local sky noise, and degree of crowding.  
However, there are three extra issues in the data that need to be addressed:
\begin{enumerate}
    \item \textit{Sharp CMD boundaries:} The first issue is already apparent in the CMDs of both Figs.\ \ref{fig:cmd} and \ref{fig:cmd_4panel}, which is that the completeness limit (lower boundary to the CMD) is marked by two distinct segments.  For bluer objects, the boundary is set by the detection limit of the redder filter, while for redder objects the boundary is set by the limit of the bluer filter.  These two boundary lines of different slopes intersect at an intermediate color where both filters have similar limits.  The transition between them is sharp.  This pair of boundary lines typically appears in any dataset where the exposure times in the two filters are well matched to reach similar overall depths.  However, this feature means that the LR color parameter $x_2$ is bilinear and will not be adequately matched by a single linear function or smooth polynomial across the entire observed range of colors.  
     \item \textit{Nonlinearity:} More generally, the simple LR model defined above assumes a logistic function $g$ that is a linear combination of all the parameters.  Just as for the color term, the other terms may not be correctly modelled by such an assumption. 
    \item \textit{Asymmetric shape:} A third issue is illustrated in Figure \ref{fig:fcompare}, drawn from the NGC 1275 data.  In the upper panel, the completeness in the outer zone ($r > 60''$) is presented in two different forms.  One is $f$ versus the measured, recovered magnitudes (black line) binned in $0.2-$mag steps:  here, as stated previously $f$ is defined as the number of stars originally put into the given magnitude bin, divided into the number of stars that were recovered in that bin, regardless of which bin those measured stars were originally placed in.  That is, bin-jumping is implicitly included.  The result is that the downturn portion of the completeness curve is asymmetric and does not match a simple sigmoid-type curve like Eq.\ \ref{eq:invlogit} very closely.
    
\end{enumerate}

\begin{figure*}
    \centering
    \includegraphics[width=0.95\textwidth]{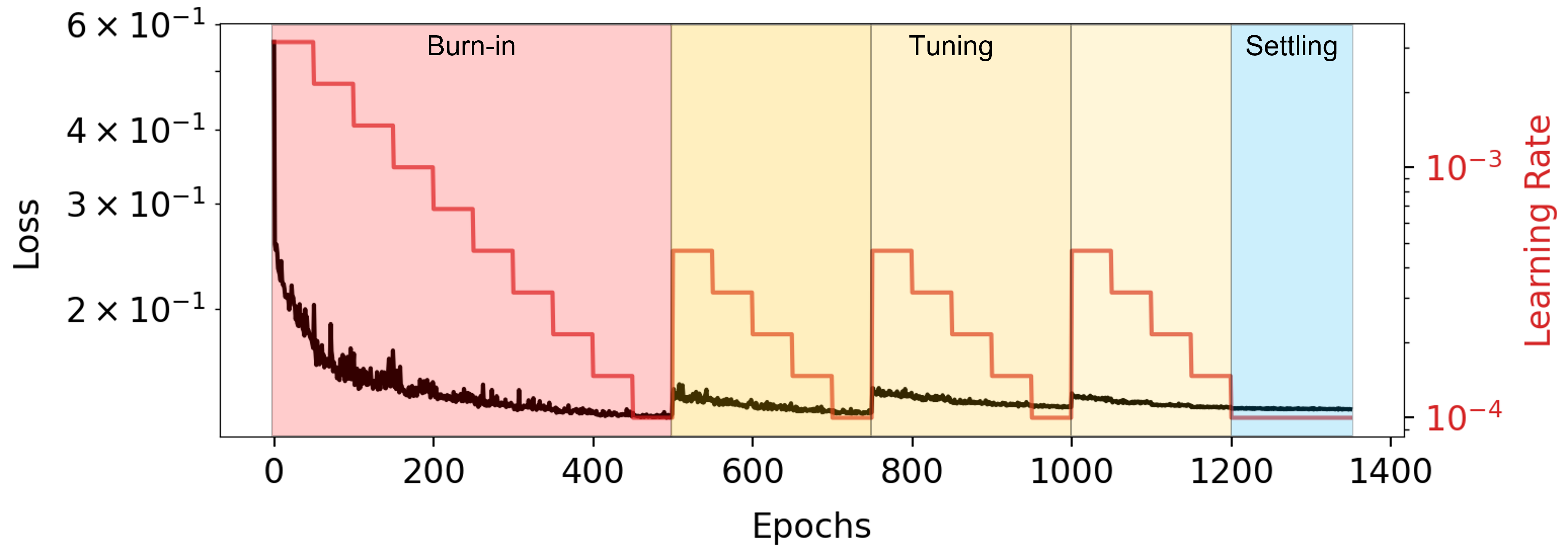}
    \caption{An example of our multi-step training scheme for our neural network (NN) model (see Section \ref{nn_final}) over artificial star data from NGC 1275. The black curve shows the average (log-)loss values over the training set as over each epoch, while the red curve shows the learning rate associated with each epoch. The behaviour for the burn-in phase, our three tuning phases, and the final settling phase are highlighted in shades of red, yellow, and blue, respectively, along with the associated amount of training data used in each phase.} %We found this ``tiered'' approach to training gave much more robust final NNs and slightly improved performance over standard training approaches.
    \label{fig:loss}
\end{figure*}

Note that the other curve in Figure \ref{fig:fcompare} is $f$ plotted versus the \emph{input} magnitude $m_{\rm in}$ (red line).  Here, $f$ is defined differently:  it is the number of stars originally put into the given bin, divided into the number of those same stars that were recovered, regardless of which bin they reappeared in.  The plot of $f$ versus input magnitude is much more nearly symmetric and matches a true sigmoid curve (dashed red line in Figure \ref{fig:fcompare}) very well.  However, as stated above, for any real data the true magnitudes are unknown and only the measured magnitudes can be used. The effects of bin-jumping are explicitly shown in the lower panel of Figure \ref{fig:fcompare}:  here, the fractional excess equals the number of recovered stars that entered the given bin \emph{from other bins} minus the number of stars that were originally in the bin but were recovered in other bins, and then  divided by the number of input stars for that bin (i.e.\ the fractional excess is the relative gain of each bin at the expense of other bins).  The large spike near F475W $\sim 28.5$ fairly near the 50\% completeness level is due to many fainter stars \emph{near} the completeness limit that are randomly scattered upward to brighter magnitudes and thus recovered.  But other such stars that were scattered downward to fainter levels are mostly unrecovered and lost (Eddington bias).  This effect distorts the shape of the $f-$curve and in particular shifts the estimated 50\% completeness point (the limiting magnitude) fainter by a quarter of a magnitude or more.  

In short, because of these effects, the LR model in its basic form will not account accurately for a more general variety of completeness curves. We show this explicitly in Figure \ref{fig:lrcompare}, where we highlight the predicted recovery probability for NGC 3377 from the best-fit simple LR model versus the original artificial star tests, binned over the CMD. The results clearly show the key issues at play: simple LR is not able to capture the sharp CMD boundaries or the asymmetric steep dropoff, and instead smoothly averages across both features (highlighted in the red box).

These results motivated us to seek an extended model capable of handling different forms of the completeness curves that do not assume symmetry or other limiting features.

\begin{figure*}
    \centering
    \includegraphics[width=0.95\textwidth]{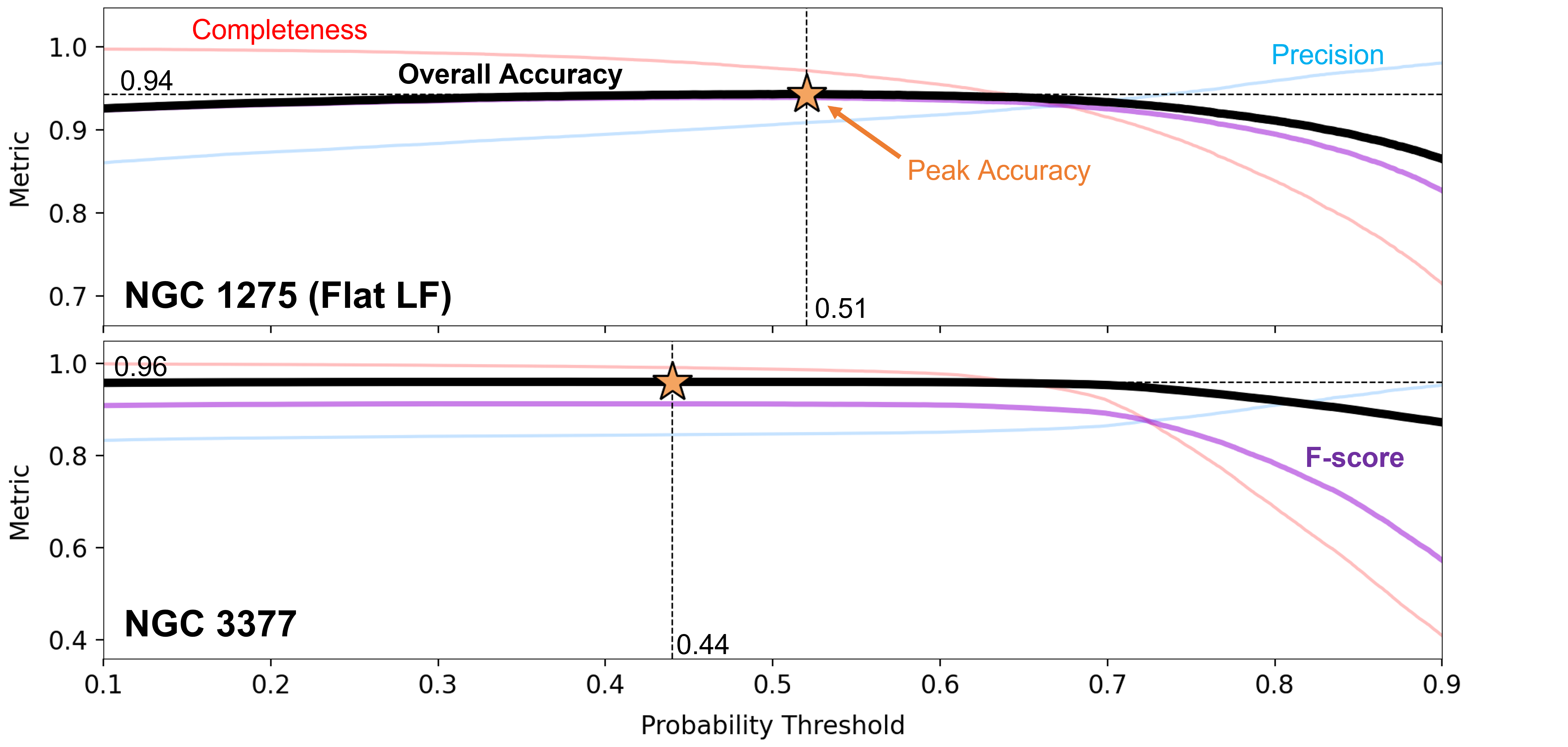}
    \caption{The accuracy (black), completeness (red), precision (blue), and F-score (purple) for the NN models trained over NGC 1275 assuming a flat LF (top) and NGC 3377 (bottom), plotted as a function of the probability threshold used to decide whether an object is successfully recovered or not. The peak accuracy values are highlighted as an orange star. Both models are able to achieve $> 94\%$ accuracy with stable performance over a wide range of values.}
    \label{fig:metrics}
\end{figure*}

\subsection{From Logistic Regression to Neural Networks} \label{nns}

We will now extend our LR model to build up the foundation of an artificial neural network (NN). In its most basic form, NNs comprise a collection of \textit{neurons}, each of which takes in a given set of inputs $\mathbf{X}$, transforms them with a set of coefficients $\bm{\gamma}$ (often referred to as \textit{weights and biases}), and then outputs the results based on an \textit{activation function} $f(x)$. In essence, each neuron performs the operation $f(x = \bm{\gamma} \mathbf{X})$, which we will write as
\begin{equation}
    {\rm Single\:Neuron}:\:\: \underbrace{\mathbf{X}}_{\rm Input} \underbrace{\overbrace{\xrightarrow[\quad \quad f(x)\quad \quad]{\bm{\gamma}}}}_{\rm Activation\:Function}^{\rm Weights\:and\:Biases} \underbrace{\chi}_{\rm Output}
\end{equation}
to emphasize that the transformation from the input data $\mathbf{X}$ to the output value $\chi$ depends on both the weights and biases $\bm{\gamma}$ and the chosen activation function $f(x)$. 

In this single neuron case, if we specifically choose our activation function to be the logistic (i.e. sigmoid) function in \eqref{eq:invlogit} and input the same data used in Section \ref{lr}, the result is \textit{exactly} equivalent to performing LR over our data and the output $\chi$ will be equivalent to the recovery probability $p$. In other words, \textit{LR can be interpreted as a NN with a single neuron using a sigmoid activation function}.

We can now use this intuition to extend LR with NNs. By combining $j=1,\dots,m$ individual neurons together, we can create a ``fully-connected'' \textit{layer} that takes in a given input $\mathbf{X}$ and returns a new collection of $m$ outputs via
\begin{equation}
    {\rm Single\:Layer}:\:\: \underbrace{\mathbf{X}}_{\rm Input} \quad \overbrace{\xrightarrow[\quad \quad f(x) \quad \quad]{\{ \bm{\gamma}_j \}_{j=1}^{m}}}^{\rm Multiple\:Neurons} \underbrace{\{\chi_j\}_{j=1}^{m}}_{\rm Multiple\:Outputs}
\end{equation}
where each $\chi_j$ is the output from each individual neuron indexed by $j$. 

Unlike the single neuron case, we can no longer interpret these outputs as a simple probability, since we ultimately are trying to predict a single number , the recovery probability $p$, rather than multiple numbers. To turn this set of new outputs $\bm{\chi} = \{\chi_j\}_{j=1}^{m}$ into a final predicted probability $p$ then, we therefore need to apply a ``probability conversion function'' $f_p(x)$. The input to this function will be a single number $x=\bm{\gamma}_p \bm{\chi}$ that we will take to be a linear combination of the input $\bm{\chi}$ values based on an additional set of coefficients $\bm{\gamma}_p$. The output of this function $f_p(x=\bm{\gamma}_p\bm{\chi}) \in [0, 1]$ will then need to be bounded between $0$ and $1$ so that it can be properly interpreted as a probability. Without loss of generality, we can consider this function to be the sigmoid function, making the connection with LR again explicit.\footnote{For completeness, we wish to note that it is straightforward to extend these results to the multi-class setting by replacing the logistic function with the \textit{softmax function}, where the probability $p_i$ of being in class $i$ out of $m$ total classes given inputs $\bm{\chi}$ and weights $\bm{\gamma}_p$ is $p_i = \exp\left({-\gamma_{p,i} \chi_i}\right) / \sum_{j=1}^{m} \exp\left({-\gamma_{p,j} \chi_j}\right)$. This reduces to the logistic function in the case where $m=2$.}

Putting this together, the $m$ neurons and their corresponding weights/biases $\bm{\Gamma} = \{ \bm{\gamma}_j\}_{j=1}^{m}$ are often together referred to as a \textit{hidden layer} (since we do not observe their outputs directly) and the total composition of the NN (the number of neurons it has, the activation functions used, etc.) is often referred to as its \textit{architecture}. In this particular case, we end up with a \textit{single-layer perception (SLP)}:
\begin{equation}
    {\rm SLP}:\:\: \mathbf{X} \underbrace{\xrightarrow[\quad \quad f(x) \quad \quad]{\bm{\Gamma}} \bm{\chi}}_{\substack{\text{Hidden Layer} \\ \text{(Feature Extraction)}}} \underbrace{\xrightarrow[\quad \quad f_p(x) \quad \quad]{\bm{\gamma}_p} p}_{\substack{\text{Probability Layer} \\ \text{(Logistic Regression)}}}
\end{equation}
where $\bm{\Gamma}$ is now a matrix whose size is equal to the number of input parameters and the number of output parameters (i.e. the number of neurons in the layer). 

In the SLP model, the final classification layer can be directly interpreted as simply performing LR over a new set of inputs $\bm{\chi}$ instead of the original inputs $\mathbf{X}$. The previous hidden layer therefore engages in what is often called \textit{feature extraction} or \textit{representation learning}, i.e. learning how to process the data in order to improve the final LR. In other words, \textit{NNs can be viewed as applying LR using learned representations of the data rather than over the data directly}. 

\begin{figure*}
    \centering
    \includegraphics[width=0.98\textwidth]{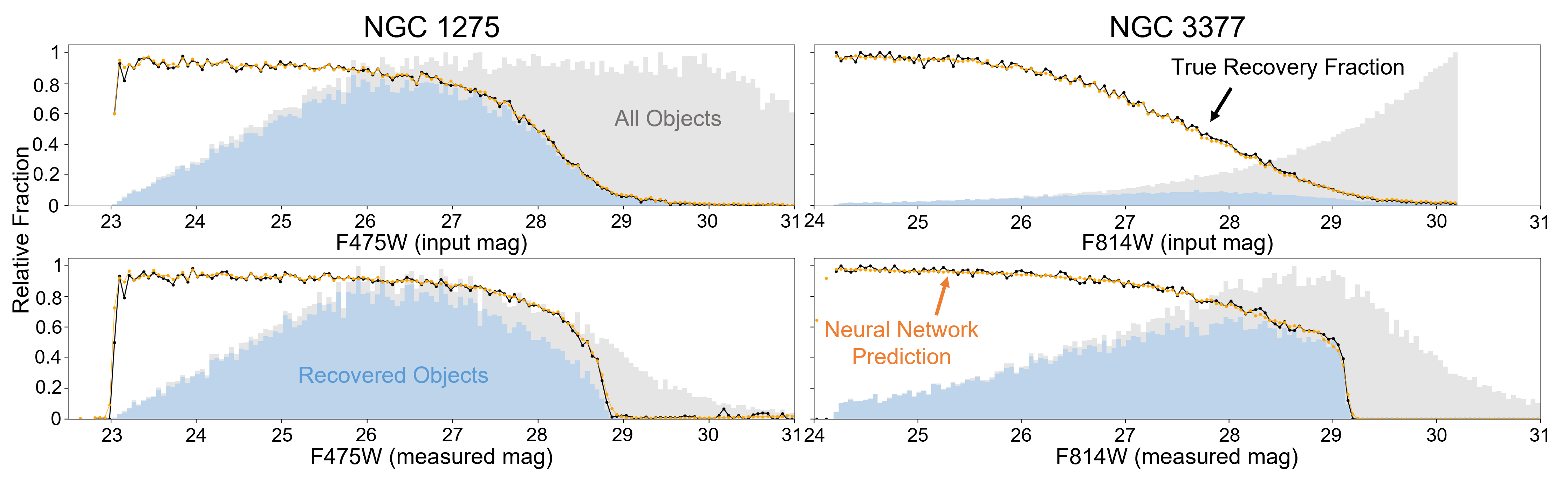}
    \caption{The predicted completeness/recovery fraction $\hat{f}(m)$ from our best-fit NN models binned in magnitude (orange line) versus the true completeness fraction $f(m)$ from the ASTs (black line) for NGC 1275 (left) and NGC 3377 (right). The input magnitude distribution versus the recovered magnitude distribution are shown for reference in light gray and blue, respectively (i.e. the black line is the ratio of the blue and gray). The top panels show the behaviour as a function of the (unknown) input magnitudes $m_{\rm in}$, which behave much like a logistic function (see Figure \ref{fig:fcompare}). The bottom panels show the behaviour over the measured (known) magnitudes, which show the NNs can reproduce the same complex behaviour highlighted in Figure \ref{fig:fcompare}.} %We see that the neural networks are able to accurately reproduce the complex dependencies in observed magnitudes and that the predicted completeness curves also agree well when plotted against the underlying true magnitudes.
    \label{fig:fmag}
\end{figure*}

Building on this intuition, we can interpret each neuron (and associated weights/biases) in the hidden layer as learning a particular feature to help interpret the data. Under the assumption that each neuron's activation function $f(x)$ is a logistic, for example, this means that the learned features are based on applications of LR over the original input data $\mathbf{X}$. In other words, \textit{we can broadly think of each neuron representing a \textit{basis function} and the outputs $\bm{\chi}$ representing the representation within that basis.} This fact also means there is no longer a requirement that $f(x)$ is a logistic function -- as long as $f_p(x)$ is logistic, $f(x)$ can be any set of arbitrary functions that can serve as an efficient basis to capture features in our data. This will be crucial to helping to resolve some of the issues with standard LR highlighted in Section \ref{lr}.

As a brief aside, we can understand this general concept of a hidden layer and activation functions as a learned basis by making a broad analogy with Fourier Series. In a Fourier series, we can write down any arbitrary function as a linear combination of sine and cosine functions. However, not all functions can be represented equally efficiently in this basis, with square waves requiring many terms to get to a specified level of accuracy. This behaviour is because localized, sharp edges are difficult to represent with a globally smooth set of basis functions (sines and cosines). Likewise, the types of functions that can be built using of the outputs of previous activation functions $\bm{\chi}$ also depends on the activation function itself. Sigmoid functions, for instance, are smooth and continuous and therefore, just like with Fourier series, serve as more efficient representations of similarly smooth functions. Other activation functions (see Section \ref{nn_final}) can serve as efficient representations for different types of complex functions.

To continue, we finally introduce the general \textit{multi-layer perception (MLP)} model, where multiple hidden layers are present before the final probability conversion layer:
\begin{equation}
    {\rm MLP}:\:\: \mathbf{X} \underbrace{\xrightarrow[f^{[1]}(x)]{\bm{\Gamma}^{[1]}} \bm{\chi}^{[1]} \xrightarrow[f^{[2]}(x)]{\bm{\Gamma}^{[2]}} \bm{\chi}^{[2]} \cdots}_{\substack{\text{Multiple Hidden Layers} \\ \text{(Feature Extraction)}}} \underbrace{\xrightarrow[\quad \quad f_p(x) \quad \quad]{\bm{\gamma}_p} p}_{\substack{\text{Probability Layer} \\ \text{(Logistic Regression)}}}
\end{equation}
Using our intuition from the simpler SLP model with one hidden layer, we can interpret our MLP model as simply having a more involved feature extraction process where, at each layer, a set of basis function representations are learned, those are applied to the input data, and the outputs are passed on to the next layer (where the process is repeated).

\begin{figure*}
    \centering
    \includegraphics[width=0.98\textwidth]{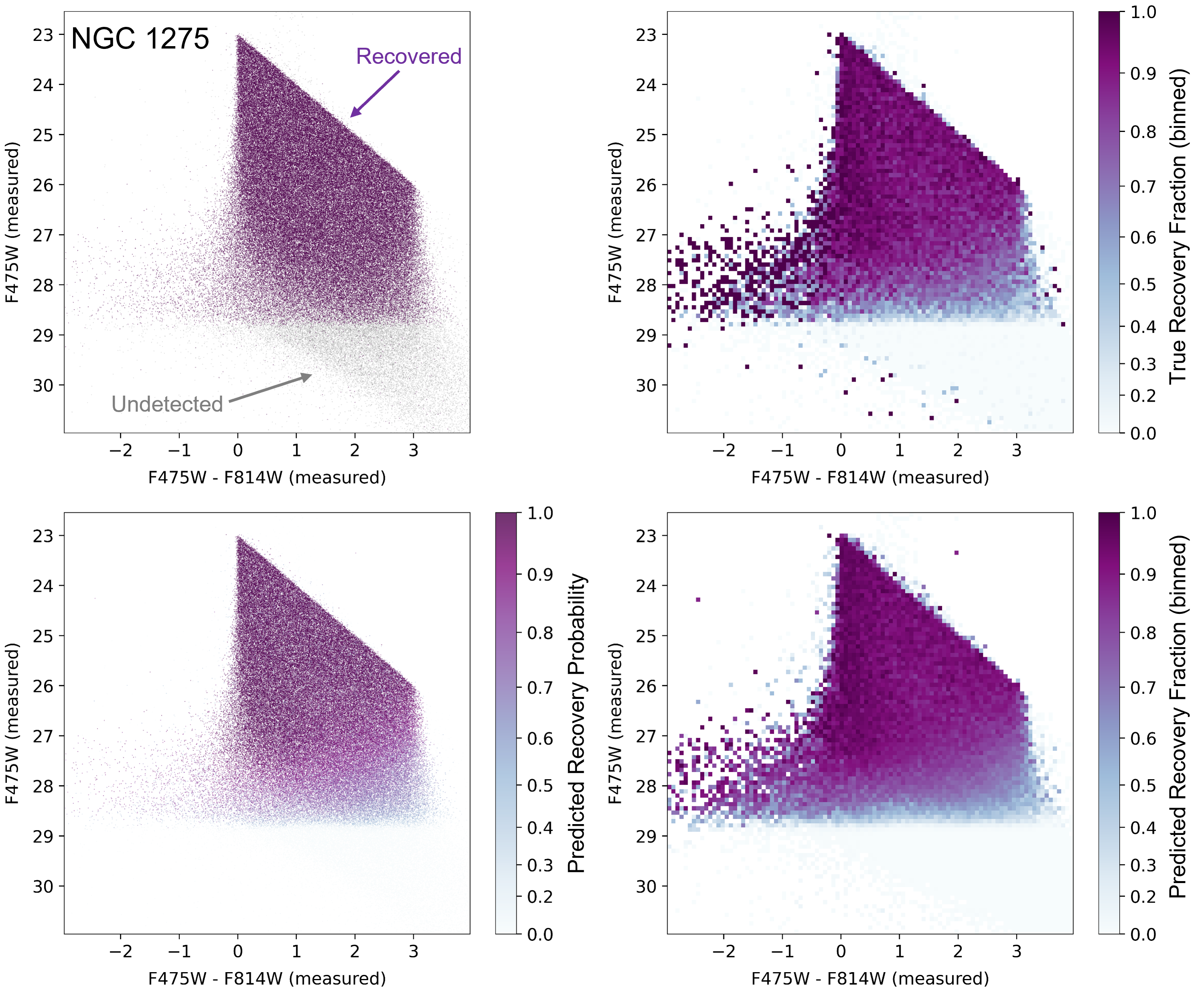}
    \caption{The dependence of stellar recovery over the measured CMD from the NGC 1275 ASTs. \textit{Top left:} The individual input artificial stars classified by whether they were recovered (purple) or not (gray). \textit{Top right:} The recovery fraction estimated in bins across the CMD (note that the colour scale is non-linear to emphasize the behaviour closer to 1). \textit{Bottom left}: As the top left panel, but now coloured by the predicted recovery probability from our NN model. \textit{Bottom right}: As the top right panel, but now coloured by the predicted recovery fraction from our NN model. We find excellent agreement across the CMD, with the top/bottom right panels providing strong evidence that the NN predictions are able to smoothly interpolate over the true recovery fraction even in the presence of shot noise.}
    \label{fig:cmd_ngc1275_ast}
\end{figure*}

Based on the above derivations, the main connections between LR and NNs that motivate our final model can be summarized as follows:
\begin{enumerate}
    \item Both methods (eventually) apply a logistic function (with learned coefficients) to convert inputs into output probabilities. Indeed, a NN with a single neuron using a logistic activation function is identical to LR.
    \item The hidden layers that are present in a NN prior to the final probability conversion (logistic regression) layer allow the NN to learn features from the data in order to improve the final LR operation. More hidden layers allows for learning potentially more complex and/or non-linear features.
    \item The types of features NNs can learn at each layer are based on the activation functions used for the neurons in that layer. In other words, each layer can be broadly interpreted as a basis function expansion and the individual neurons as individual basis functions.
\end{enumerate}

\subsection{Defining Our Neural Network Model} \label{nn_final}

With the motivation from Section \ref{nns} in hand, we will now proceed to define the exact NN-based model used in this work. After some testing, we found that the best-performing NN was a simple, fully-connected network with three layers, each with 200 neurons. Due to the sharp and asymmetric recovery boundaries, we found that the overly smooth basis functions such as the logistic or tanh were unable to capture the sharp transitions in the CMD very efficiently. As a result, we opted to use the \textit{rectified linear unit (ReLU)}
\begin{equation}
    {\rm ReLU}(x=\bm{\beta}_i \mathbf{X}) = \max\left(0, \bm{\beta}_i \mathbf{X}\right)
\end{equation}
as the activation functions across all three layers, which is linear when $\bm{\beta}_i \mathbf{X} > 0$ and $0$ otherwise. This allows the hidden layers of our NN to model the data using the subset of all piece-wise linear functions, which is able to better accommodate rapid local changes in the recovery probability across the CMD seen in Figure \ref{fig:lrcompare}.

\begin{figure*}
    \centering
    \includegraphics[width=0.98\textwidth]{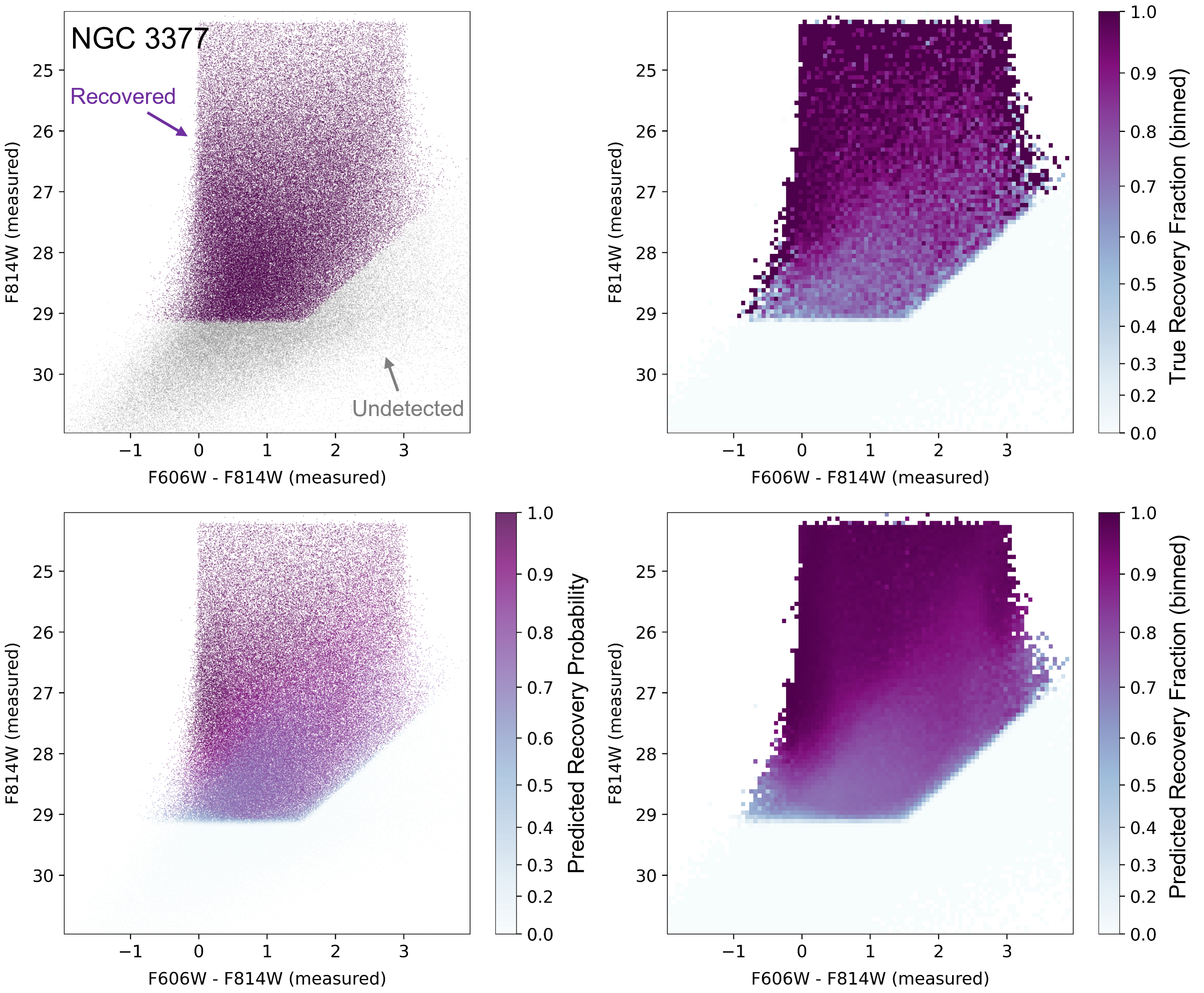}
    \caption{As Figure \ref{fig:cmd_ngc1275_ast}, but now for NGC 3377. We see the same trends but even stronger signs that the NN model is able to derive accurate estimates of the recovery fraction even in the presence of shot noise.}
    \label{fig:cmd_ngc3377_ast}
\end{figure*}

We train our neural network using \texttt{scikit-learn} \citep{scikit-learn} to optimize the classification log-likelihood $\ln \mathcal{L}(\bm{\theta})$ (i.e. minimize the log-loss $LL(\bm{\theta})$) via
\begin{equation}
    \hat{\bm{\theta}} = \argmin_{\bm \theta} \lbrace LL(\bm{\theta}) \rbrace = \argmax_{\bm\theta} \lbrace \ln \mathcal{L}(\bm{\theta})\rbrace
\end{equation}
where the log-likelihood is defined as
\begin{equation}
    \ln \mathcal{L}(\bm{\theta}) = \sum_{i=1}^{n} y_i \ln(p(\mathbf{X}_i;\bm{\theta})) + (1 - y_i) \ln(1 - p(\mathbf{X}_i;\bm{\theta}))
\end{equation}
and where $y_i={1,0}$ states whether the $i$th object was recovered ($1$) or not ($0$), $p(\mathbf{X}_i;\bm{\theta})$ is the recovery probability predicted from the NN given the weights/biases $\bm{\theta}$, and the sum over $i$ is taken over all objects $n$ in a given training batch. We opt to train using 80\% of the AST data (with the remaining 20\% reserved for validating/testing performance) with a batch size of $n=256$. We apply a small L2 regularization of $\alpha=10^{-4}$ to prefer slightly sparser distributions of weights/biases. We perform the optimization itself using the \texttt{adam} optimizer \citep{adam}, which uses a combination of stochastic gradient descent (SGD) and momentum-based strategies.

\begin{figure*}
    \centering
    \includegraphics[width=0.85\textwidth]{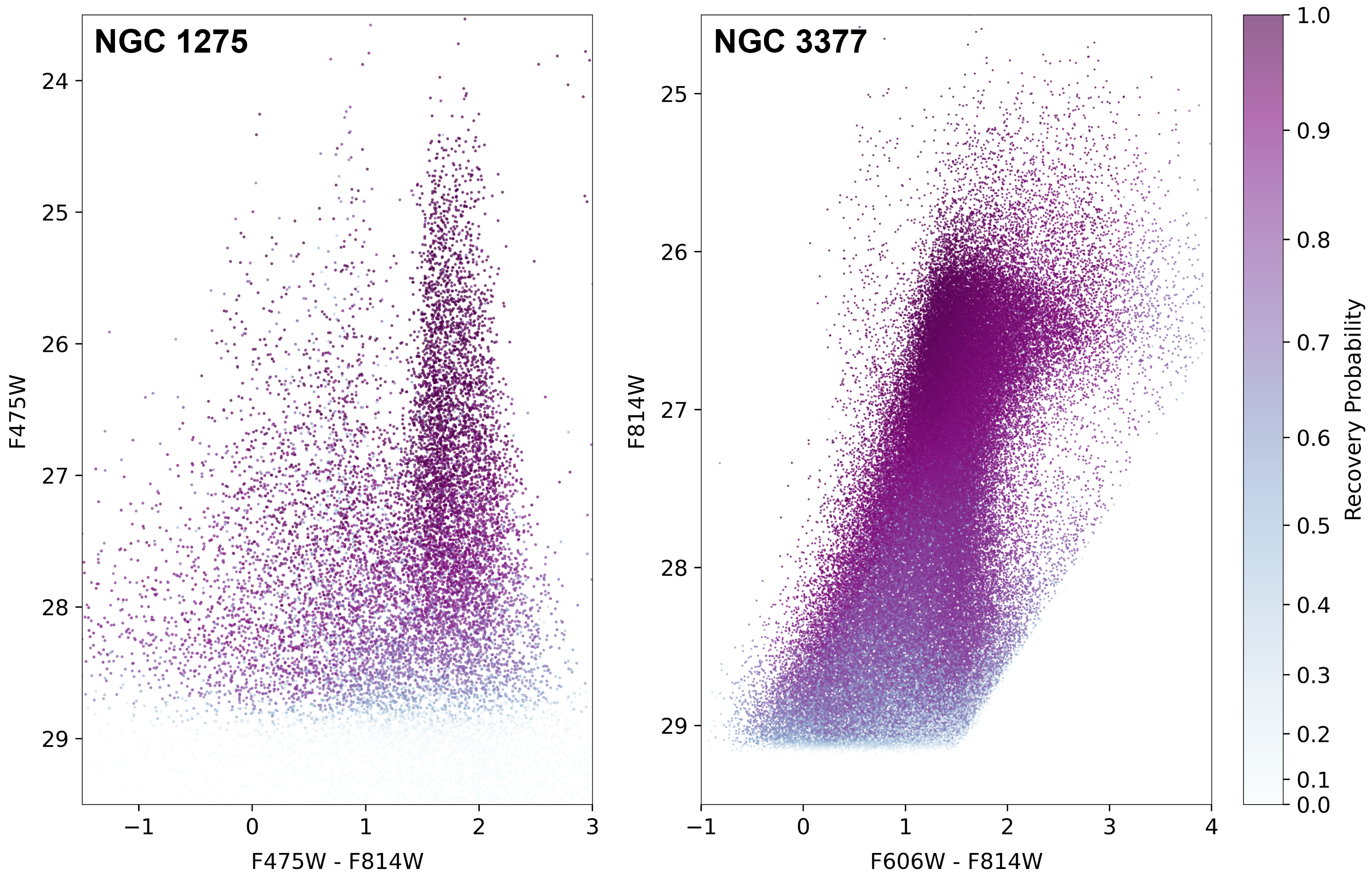}
    \caption{The measured CMDs for NGC 1275 (left) and NGC 3377 (right) with individual objects coloured by the predicted recovery probability from our best-fit NN models.  As in the previous graphs, note the nonlinear color scale.}
    \label{fig:cmd_combined}
\end{figure*}

We found that getting the NN model to reproduce the exact recovery fraction across both magnitude and color to high precision required a stepped training procedure involving a \textit{learning rate schedule} (i.e. time-dependent learning rate) accompanied by different amounts of training data. The learning rate $a(t)$ at a given iteration multiplies the gradient, reducing the overall step taken after processing a particular training batch. We found a multi-step approach where adjusting the learning rate along with the quantity of input data helped the network learn the correct ``rough behaviour'' first, followed by slow adjustment period to properly reproduce all the features seen in the model data. We confirmed across a number of ASTs, training data sizes, and random seeds that this approach helped guarantee robust final solutions. 

Our final approach involved three main phases:
\begin{enumerate}
    \item \textit{Burn-in phase}: In the first ``burn-in'' phase, we used 5\% of the training data accompanied by a learning rate schedule with 10 values logarithmically spaced from $10^{-2.5}$ to $10^{-4}$. We train for 50 epochs (i.e. full passes through a random reshuffling of the input training data) for each value of the learning rate. This phase helps the NN establish a good approximation to the functional form of the recovery probability.
    \item \textit{Tuning phase}: In the second ``tuning'' phase, we used 10\% of the training data accompanied by a learning rate schedule with 5 values logarithmically spaced from $10^{-3.33}$ to $10^{-4}$ and again training for 50 epochs each. This entire process was then repeated using 20\% of the training data, and then finally with 100\% of the training data. This phase allows the NN to appropriately improve its predictions based on larger segments of the training data
    \item \textit{Settling phase}: In the final ``settling'' phase, we used 100\% of the training data with a constant learning rate of $10^{-4}$ and ran for 100 epoch. This process helps ensure the final network parameters are fine-tuned and remain stable.
\end{enumerate}
The behaviour of the loss function over time for one of the networks trained over the NGC 1275 artificial-star data is shown in Figure \ref{fig:loss}, highlighting the varying loss rates and overall performance over the training set. The exact NN models we trained\footnote{Note again that the model for NGC 1275 uses separate parameters for sky noise and crowding while the model NGC 3377 uses a single parameter for both (see Section \ref{model_params} for additional details).} along with the entire collection of ASTs can be found on Zenodo at \url{https://doi.org/10.5281/zenodo.8306488}.

\begin{figure*}
    \centering
    \includegraphics[width=0.98\textwidth]{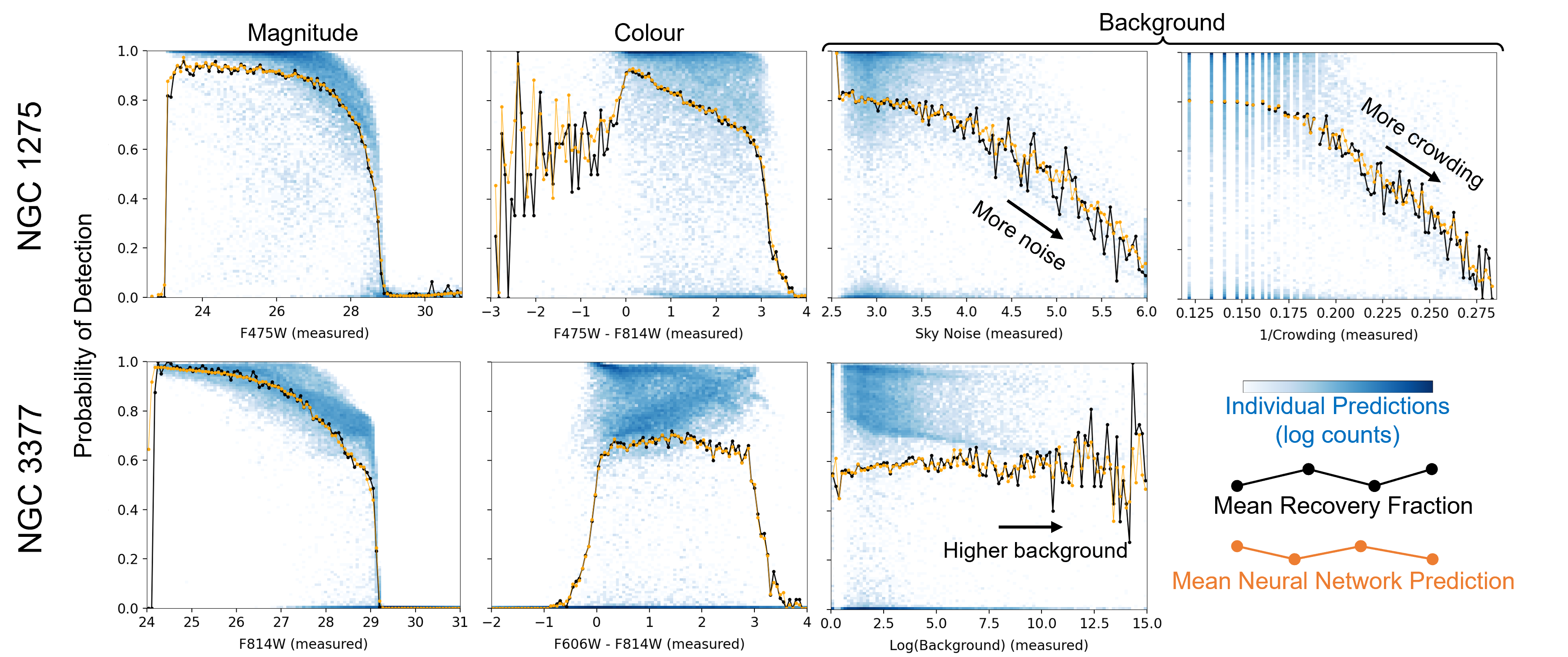}
    \caption{Similar to Figure \ref{fig:fmag}, but now plotting the predicted recovery fraction (orange) and the true recovery fraction (black) against all input parameters for NGC 1275 (top) and NGC 3377 (bottom), with the density of individual predictions shown in blue tones. This includes magnitude (far left), colour (center left), sky noise (top center right), crowding (top far right), and background density (bottom right). These individual panels highlight that our NN model predictions behave sensibly in not only magnitude (declining steeply near the detection limit) but also colour (declining outside of the main colour range) and sky noise/crowding (declining in noisier and more crowded regions). They also highlight that mean trends can mask a lot of complex behaviour that the NN model is able to successfully capture in the predicted individual recovery probabilities.  {Note that in the upper right panel, the vertical streaks of points are due to the quantized nature of the crowding variable for NGC 1275.}} %This includes correctly predicting some low recovery probabilities at bright magnitudes for NGC 1275 due to crowding near the galaxy center (top left; see also Figure \ref{fig:cmd_4panel}) as well as very noticeable group trends as a function of colour and background in NGC 3377 that almost perfectly cancel out to leave the mean behaviour unchanged (bottom left center and bottom right).
    \label{fig:prob_combined}
\end{figure*}

\section{Application to NGC 1275 and NGC 3377} \label{nn_ngc}

\subsection{Model Performance}

We first show the general performance of our NN model as a function of the predicted recovery probability $\hat{p}_i = p(\mathbf{X}_i;\hat{\boldsymbol{\theta}})$ for each object $i$, where we classify an object as recovered ($\hat{y}_i=1$) or unrecovered/lost ($\hat{y}_i=0$) following
\begin{equation}
    \begin{cases}
        \hat{y}_i = 1 \:\:({\rm recovered}) & {\rm if}\:\: \hat{p}_i \geq p_{\rm thresh} \\
        \hat{y}_i = 0 \:\:({\rm lost}) & {\rm if}\:\: \hat{p}_i < p_{\rm thresh}
    \end{cases}
\end{equation}
Here $p_{\rm thresh}$ is the minimum probability threshold for an object to be considered ``recovered''. Based on this, we consider four standard primary metrics that gauge how often our prediction $\hat{y}_i$ compares to the truth value $y_i$:
\begin{align}
    {\rm Accuracy} &= \frac{\sum_i \mathbbm{1}(y_i=\hat{y}_i)}{\sum_i 1} = \frac{n_{\rm corr}}{n} \label{eq:acc} \\
    {\rm Completeness} &= \frac{\sum_i \mathbbm{1}(\hat{y}_i=y_i=1)}{\sum_i \mathbbm{1}(y_i=1)} = \frac{n_{\rm rec,corr}}{n_{\rm rec,true}} \label{eq:comp} \\
    {\rm Precision} &= \frac{\sum_i \mathbbm{1}(\hat{y}_i=y_i=1)}{\sum_i \mathbbm{1}(\hat{y}_i=1)} = \frac{n_{\rm rec,corr}}{n_{\rm rec,pred}}  \label{eq:prec} \\
    {\rm F\text{-}score} &= \left[\frac{1}{2}\left(\frac{1}{{\rm Completeness}} + \frac{1}{{\rm Precision}}\right)\right]^{-1} \label{eq:fscore}
\end{align}
where $\mathbbm{1}(\cdot)$ is the indicator function which returns $1$ when the condition is true and $0$ if it is false. In other words:
\begin{itemize}
    \item the accuracy is the total fraction of correct predictions for recovered/non-recovered across the entire sample,
    \item the completeness is the fraction of objects we successfully (i.e. ``correctly'') recover out of all true objects,
    \item the precision is the fraction of objects we successfully recover out of all claimed/predicted objects, and
    \item the F-score is the harmonic mean of the completeness and the precision.
\end{itemize}

We plot these metrics for both NGC 1275 and NGC 3377 given our best-fit NN model in Figure \ref{fig:metrics}.  These results show accuracies for both galaxies that reach $>$\,94\%, which achieve several percentage points higher than the best-performing LR models. We find the accuracy is maximized for values of $p_{\rm thresh} \approx 0.5$.
%, which is theoretically expected given our chosen (log-)loss function. 
Our model also tends to be more complete than precise at these values, highlighting the robustness of our recovery (i.e. when forced to classify individual stellar predictions, the NNs generally produce 5\% to 10\% false recovered objects while only ``missing'' 2\% to 5\% of true recovered objects).

The magnitude dependence of our best-performing NN models for NGC 1275 (under the ASTs produced under the flat LF)\footnote{We find that the behaviour of the NN model for NGC 1275 under the flat LF is \textit{slightly} improved compared to the steep LF, mainly because the large number of faint, non-recovered sources (with virtually no chance of every being recovered) in the latter case skews the model towards slightly higher incompleteness estimates in order to more accurately capture this behaviour. None of the results presented in this Section change if this other NN model is used instead.} and NGC 3377 is shown in Figure \ref{fig:fmag}. We find that the NNs are able to reproduce very precisely the complex magnitude dependencies for both sets of data. Furthermore, we find that the NN predictions, when plotted against the underlying true (input) magnitudes, also correctly reproduce the overall logistic/sigmoid shape seen in Figure \ref{fig:fcompare}.

\begin{figure}
    \centering
    \includegraphics[width=.48\textwidth]{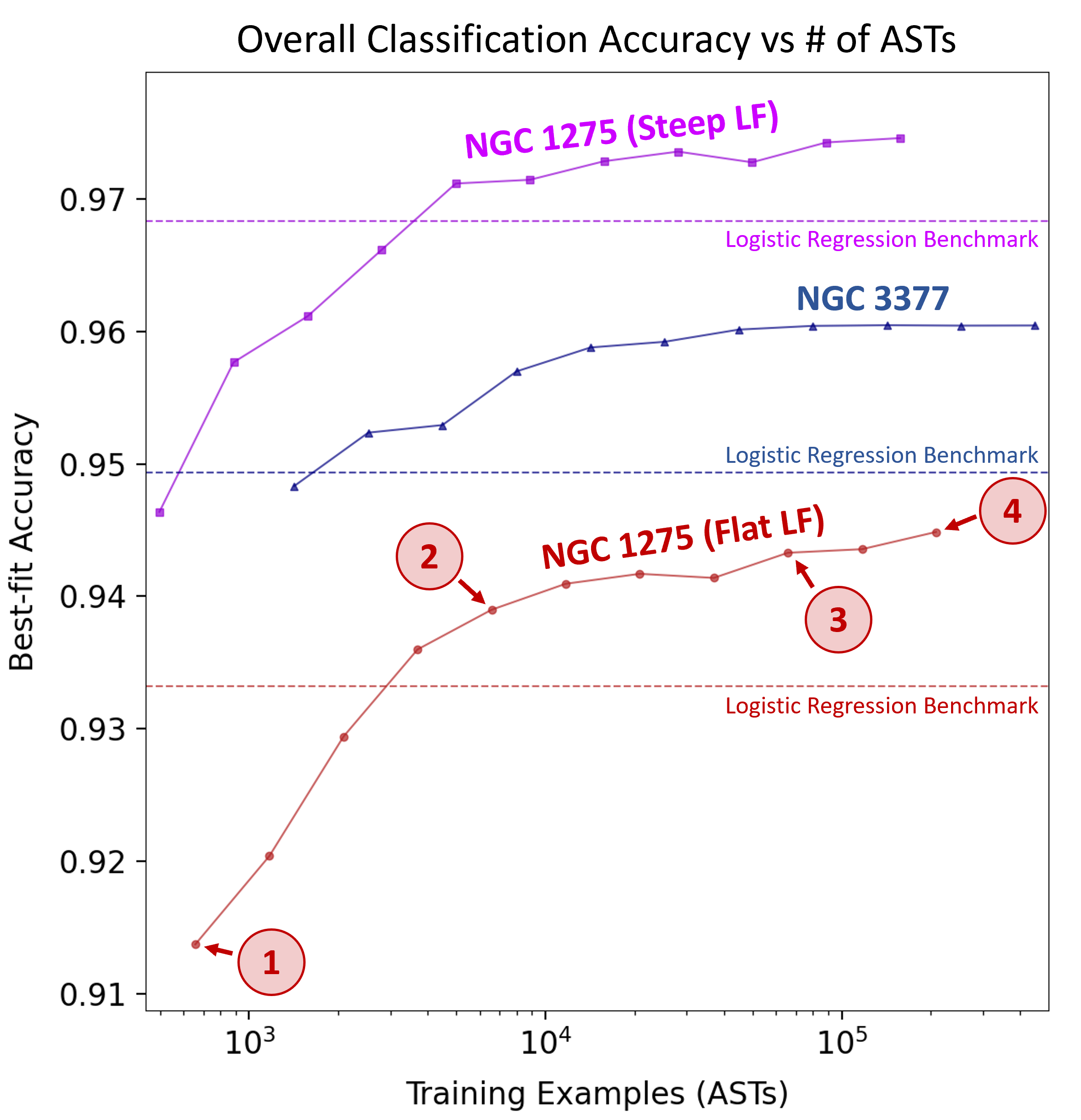}
    \caption{The change in the classification accuracy (recovered vs lost with a decision boundary of $\hat{p}(\mathbf{X}) = 0.5$) for our NN models based on the size of the training set for the three AST scenarios explored in this work: ASTs for NGC 1275 drawn from a flat LF (red), for NGC 1275 drawn from a steep LF (purple), and for NGC 3377 drawn from a steep but truncated LF (blue). The accuracy for each scenario from a simple logistic regression model (as shown in Figure \ref{fig:lrcompare}) is highlighted as horizontal dashed lines. The behavior of the four specific NN models shown in Figure \ref{fig:ast_dependency_mag} (below) is also highlighted. We find that only a few tens of thousands of artificial star tests are needed to capture the majority of the behaviour seen in the data, although additional ASTs often still help to improve performance.}
    \label{fig:ast_dependency}
\end{figure}

Finally, we highlight the behaviour across the measured CMDs from the ASTs for NGC 1275 and NGC 3377 in Figures \ref{fig:cmd_ngc1275_ast} and \ref{fig:cmd_ngc3377_ast}. We see that the NN models are able to accurately reproduce all of the features that LR struggled with in Figure \ref{fig:lrcompare}:  the strong asymmetric features, sharp, broken boundaries, and additional structure even at brighter magnitudes. Together, these demonstrate that our NN-based approach succeeds where our simple LR failed, and serves as a qualitatively and quantitatively accurate model of the associated selection function.

%\subsection{Real Data}

Based on the results above, in Figure \ref{fig:cmd_combined} we return to the real photometry for the two fields studied here. The measured CMDs for NGC 1275 (Figure \ref{fig:cmd}) and NGC 3377 (Figure \ref{fig:cmd_4panel})  are replotted, but now with each object colored by the predicted recovery probability $\hat{p}(\mathbf{X})$. 
What the complete NN model fit has allowed us to do here is to assign a reasonable recovery probability \emph{to each individual object} in the CMD, depending on its magnitude, color, location in the image field, and local crowding level.  All issues arising from binning of the data are avoided.

\subsection{Reproducing the Completeness Curve}

One of the biggest concerns surrounding the use of a more complex model such as NN is that the ``black box'' nature of the model can substantially hinder interpretability in the output predictions. While we have provided some theoretical arguments illustrating that our NN builds on more interpretable foundations such as LR, now we want to investigate whether the model is behaving appropriately under a variety of conditions.

\begin{figure*}
    \centering
    \includegraphics[width=0.98\textwidth]{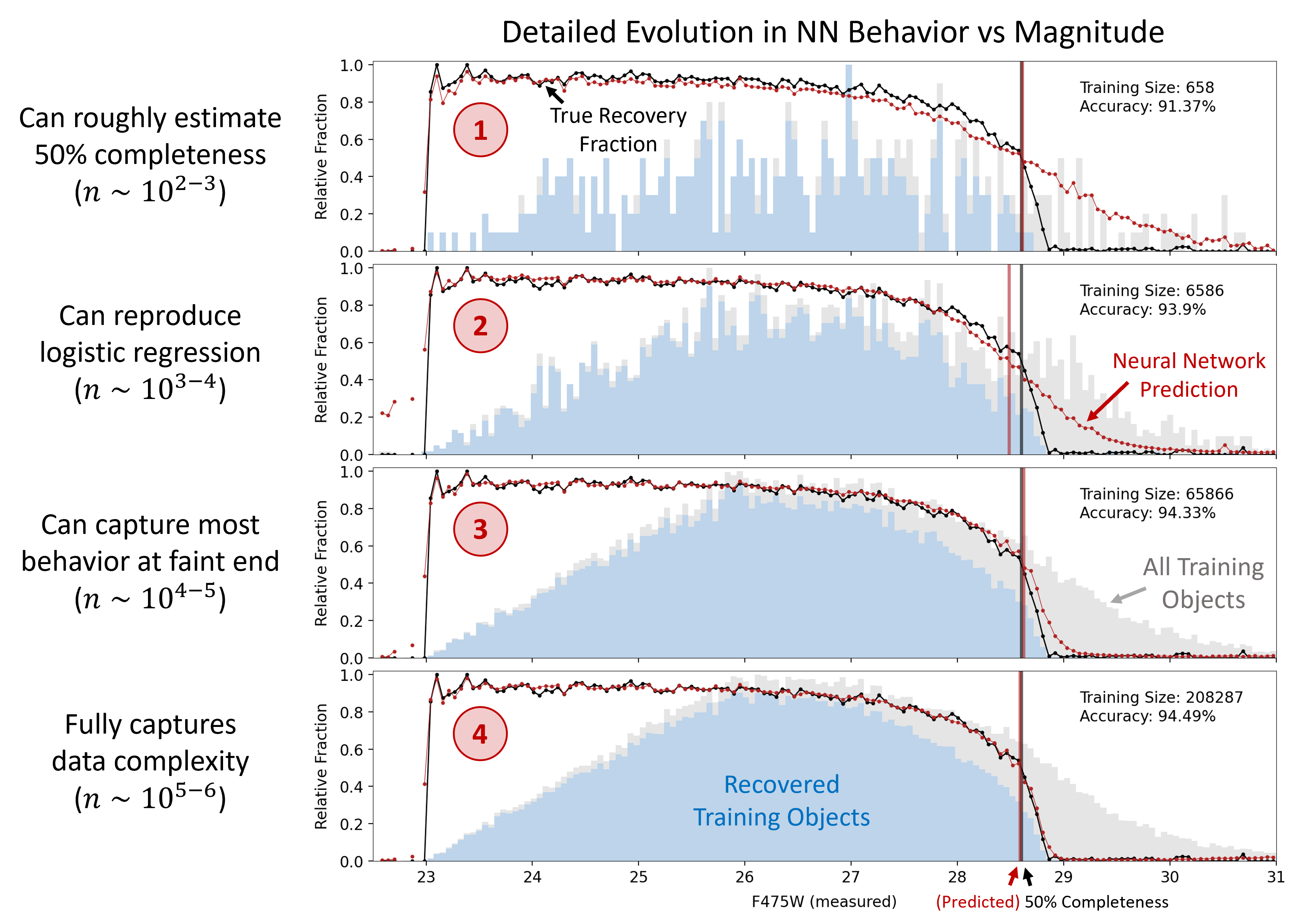}
    \caption{Similar to Figure \ref{fig:fmag}, but now highlighting the behaviour of the four NN models labeled in Figure \ref{fig:ast_dependency} and with the magnitude distribution of the training (rather than testing) data shown. The 50\% recovery limit in the testing set from the prediction and the data are also plotted as red and black vertical lines, respectively. These results help to illustrate how the NN is able to learn from increasing numbers of ASTs, and how many ASTs might be needed for various tasks.}
    \label{fig:ast_dependency_mag}
\end{figure*}

In Figures \ref{fig:fmag}, \ref{fig:cmd_ngc1275_ast}, and \ref{fig:cmd_ngc3377_ast}, we showed that the NN model predictions agree extremely well with the true recovery fraction as a function of magnitude and color. However, these plots might be hiding other troubling behavior in individual predicted probabilities, and also do not display very clearly the impact of crowding and sky noise as opposed to magnitude and color.

To address these questions, we also investigated the behavior of our NN predictions for the star-by-star probabilities $\hat{p}(\mathbf{X})$ and the average recovery fraction $\hat{f}(\mathbf{X})$ across all of our inputs. The results are shown in the panels of Figure \ref{fig:prob_combined}, which demonstrate that the NN models are able to reproduce the binned average recovery fractions across \textit{all} parameters extremely well.  The figures also demonstrate the extent to which the NNs are able to learn complex dependencies that can be easily overlooked when only binned averages are used. For instance, in the NGC 1275 field the model is able to successfully reproduce the slight incompleteness even at bright magnitudes that is caused by the very high background sky noise in the central bulge of the galaxy (see Figure \ref{fig:fcurve}).  This effect shows up as a low-probability cloud in the magnitude-based plot (upper left panel of Fig.~\ref{fig:prob_combined}).   Similarly, for the NGC 3377 field, we instead see that the NN is able to reproduce a complex distribution for individual recovery probabilities across magnitude, colour, and background density, where two population trends emerge from the more complex selection function across the CMD (see Figure \ref{fig:lrcompare}). Most of these complexities end up being completely masked if we look, e.g., only at the mean behaviour as a function of magnitude $\hat{f}(m)$ instead of the distribution of $\hat{p}(\mathbf{X})$.

\subsection{Artificial-Star Testing:  How Much is Enough?}

We now want to use our model to discuss a general and very practical question that is rarely addressed in previous literature:  how much is enough?  That is, how many ASTs do we need to ensure we can accurately characterize the recovery probability?  To investigate this question, we train new NN models with widely differing numbers of input artificial stars for the three sets of ASTs discussed in Sections \ref{ngc1275} and \ref{ngc3377}: ASTs for NGC 1275 drawn from a flat LF, ASTs for NGC 1275 drawn from a steep, rising LF, and ASTs for NGC 3377 drawn from a steep but truncated LF. To ensure stable results, each model was trained with the exact same procedure as described in Section \ref{nn_final}\footnote{The only exception to this step was the batch size was scaled down to be $<5\%$ of the total size of the training set, with a maximum allowed batch size of $n=256$ and a minimum allowed batch size of $n=32$.} and with training sets that were explicit subsets of each other, i.e. the training sets were not shuffled between objects to ensure that $\mathrm{(Training\:Set\:1)} \in \mathrm{(Training\:Set\:2)} \in \dots \in \mathrm{(Original\:Training\:Set)}$.

Our results are summarized in Figure \ref{fig:ast_dependency}, where the overall classification accuracy as defined in \eqref{eq:acc} provides a convenient way to evaluate the effectiveness of each test run.  Accuracy improves with larger numbers of input stars, but the rapid gains that occur in the small-$N$ regime where $N \lesssim 10^4$ input stars then start to saturate and level off for larger $N$.  Overall, we require only a few thousand ASTs for our NN model to exceed the performance of our initial LR benchmarks (see Section \ref{lr} and Figure \ref{fig:lrcompare}). The exact number depends on how complex the simulated recovery behaviour is: the more complex the behaviour, the fewer ASTs are needed to get substantial performance gains and to outperform simple LR. 

To understand these results more fully, in Figure \ref{fig:ast_dependency_mag} we show the estimated and true recovery fractions for a subset of the NN models trained over the NGC 1275 ASTs (flat input LF). With an input sample of only a few hundred stars to train over (top panel), we find that although the performance is poor (well below the LR benchmark), the NN is still able to estimate the 50\% recovery fraction limit relatively accurately. As we increase our training set to a few thousand stars (top middle panel), we find the NN broadly learns to mimic the behaviour of a simple LR model, such as the one shown in Figure \ref{fig:lrcompare}. Once we start using tens of thousands of stars  (bottom middle panel), we find the NN is successfully able to reproduce almost all major features seen in the data.  Finally, once we move into hundreds of thousands of stars (bottom panel), we find the NN is able to reproduce the full complexity of the input ASTs. 

While these results have been derived with our NN-based approach, they broadly imply that the size of ASTs one needs depends on how precisely one wants to estimate the recovery fraction. If it is enough just to estimate the 50\% recovery limit to within 0.2 mag or so, then only a few thousand input stars might be enough. If one wants to account more accurately for populations all the way down to (or even below) the 20\% recovery level, however, then likely $N > 10^5$  artificial stars will be required.

In addition, we have found that it is not necessary to overload the AST population with stars that are far fainter than the completeness limit where they have no chance of recovery.   Large numbers of such stars will have the unwanted effect of pulling the solutions away from the more important goal of matching the downturn region of the completeness curve.  Furthermore, if too many faint unresolved stars are added all at once to the image, they can even change the intrinsic level of sky noise.

\subsection{Handling Multiple Filters:  Extending the Model}

{Many photometric studies employ just two filters, yielding a magnitude and one color index.  This is the case our current discussion focusses on.  Magnitude and color are logical choices as input variables to a LR or NN model because they represent two astrophysically different quantities, i.e. stellar luminosity and effective temperature.  But a natural question is how the NN model could be extended to photometric studies with three or more filters, such as the SDSS survey in $ugriz$ \citep[][and a vast series of later papers]{york+2000} as perhaps the most prominent example.}

{Multiple ($N_f \geq 3$) filters open up a wider range of possibilities for the choice of variables.  From a strict computational viewpoint, the NN algorithm is designed to be able to handle these cases, but it is worth considering carefully what is actually wanted from the solution. One option might be to replace the (magnitude+color) pair with $N_f$ magnitudes, treating them all as if they were independent. A practical issue to consider may be that the measured magnitudes from the various filters will often be strongly correlated with each other (unless their central wavelengths are so far apart that they sample widely different parts of the spectrum), so that in fact they are not adding fully independent pieces of information into the solution.  (However, although their magnitudes may be correlated, their measurement uncertainties are usually not.) Another option could be to use one magnitude (probably the one giving the deepest data) and $(N_f-1)$ color indices.  Again, the different colors will often be well correlated with each other if they sample parts of the spectrum that are not too widely separated.  Still another option might be to use just one magnitude, and just one color index from the two filters most widely separated in wavelength.}

{For any of these choices, we assume that the effective limits of the photometry are similar in all the bands, so that most of the original catalog of sources that one is interested in will have useful measurements in every filter.  If that is not the case, then the completeness of the data will be dominated by the shallowest of the filters.  In that case, careful consideration should be given to which bandpasses should actually be included in the analysis.}

{There is no ideal single solution to the multiple-bandpass situation that would clearly be best for all possible alternatives.  Above all, it is important to know the features of one's own database, to have a clear view of what kind of result is desired, and to adapt the choices of input variables accordingly.  Given that useful assessments of completeness can be obtained from surprisingly small ASTs (see above), our recommendation would be to use the preliminary testing stage to experiment with different combinations of input variables and settle on a combination that maximizes the classification accuracy.}

\section{Summary} \label{summary}

In this paper, we explore the application of a neural network to the problem of completeness for optical/NIR photometry.  As testbeds for developing the model, we use HST multicolor imaging data for the star cluster population around the Perseus giant galaxy NGC 1275, and for the RGB stellar population in the nearby early-type galaxy NGC 3377.  
The main conclusion of this study is that neural networks (NNs) provide an attractive approach by comparison with  conventional methods for measuring photometric completeness, such as logistic regression (LR) or binning as a function of magnitude and color.  

Within the NN model, the effects of the parameters that govern the recovery fraction -- magnitude, color, local sky noise, and local crowding -- can all be handled \emph{simultaneously} within a single model.  The degree of completeness can be determined for objects located anywhere in the field, and at any location on the color-magnitude diagram, from a single functional form.  The solutions have the additional advantage that they clearly reveal the relative importance of each parameter in the given situation.

{It will be apparent to the reader that mapping out the recovery fraction of the photometry as a function of all the variables (magnitude, color, local sky and crowding) is essential for several purposes in later data analysis.  For example, comparing the luminosity function of the objects in different places within the field, or more generally their distribution in the CMD or color/color diagram, must account for the way the completeness changes across the distribution.  Deducing the range of underlying source parameters such as mass, metallicity, age, or reddening follows as well.}

Artificial-star tests continue to be the key to revealing the pattern of recovery completeness.  From the experiences gained in this study, we recommend the following approach to running ASTs:
\begin{enumerate}
    \item Do a quick initial AST run of $\sim 10^3$ input stars, enough to find the key magnitude range over which the completeness drops from one to zero;
    \item Carry out a second major run of $\sim 10^{4-5}$ stars that is more carefully designed to cover the range of magnitudes, color, sky noise, and crowding that are present in the real images.
\end{enumerate}

The NN code and the full AST data used in this study can be found at Zenodo at \url{https://doi.org/10.5281/zenodo.8306488} under \dataset[DOI: 10.5281/zenodo.8306488]{https://doi.org/10.5281/zenodo.8306488}.  The code is named COINTOSS (COmpleteness wIth Neural networks TO Simulate Stellar recovery).
The HST data presented in this article were obtained from the Mikulski Archive for Space Telescopes (MAST) at the Space Telescope Science Institute. The specific observations analyzed can be accessed via \dataset[DOI: 10.17909/wb05-2618]{https://doi.org/10.17909/wb05-2618} and \dataset[DOI: 10.17909/kxhv-aj70]{https://doi.org/10.17909/kxhv-aj70}.

\section*{Acknowledgements}

WEH led the original data collection, reduction, analysis, and the ASTs, and drafted the majority of Sections 1-5 and 8. JSS led the completeness curve modelling, drafted the majority of Sections 6-7, and provided contributions to the remainder of the text.
The authors thank John Blakeslee for helpful conversations. JSS would also like to thank Rebecca Bleich for reminding him ``you don't do AI''.
The authors acknowledge the financial support of NSERC (Natural Sciences and Engineering Research Council of Canada).
\facility{HST (ACS)}
\software{\texttt{DOLPHOT} \citep{dolphin2000}, \texttt{DAOPHOT} \citep{stetson1987}, \texttt{AstroDrizzle} \citep{hack+2012}}, \texttt{scikit-learn} \citep{scikit-learn}.

\bibliographystyle{apj}
\bibliography{completeness}

%\include{alldata}

%\makeatletter\@chicagotrue\makeatother

\label{lastpage}

\end{document}